\documentclass[aps, pra, twocolumn, amsmath, amssymb, showpacs, nofootinbib]{revtex4-1}

\usepackage[utf8]{inputenc}
\usepackage{graphicx}
\usepackage{units}
\usepackage{color}
\usepackage[pdftex, colorlinks=true, linkcolor=myblue, citecolor=myblue,
urlcolor=myblue]{hyperref}
\usepackage{times}

\definecolor{myblue}{rgb}{0,0,1}

\newcommand{\strasbourg}{Universit\'e de Strasbourg, CNRS, Institut de Physique et Chimie des Mat\'eriaux de Strasbourg,
UMR 7504, F-67000 Strasbourg, France}

\begin{document}

\title{Orbital magnetism in ensembles of gold nanoparticles}

\author{Mauricio G\'omez Viloria}
\affiliation{\strasbourg}

\author{Guillaume Weick}
\email{guillaume.weick@ipcms.unistra.fr}
\affiliation{\strasbourg}

\author{Dietmar Weinmann}
\affiliation{\strasbourg}

\author{Rodolfo A.\ Jalabert}
\affiliation{\strasbourg}


\begin{abstract}
The last two decades have witnessed various experiments reporting very 
unusual magnetic properties of ensembles of gold nanoparticles 
surrounded by organic ligands, including ferromagnetic, paramagnetic, 
or (large) diamagnetic responses.
These behaviors are at odds with the small diamagnetic response of 
macroscopic gold samples.
Here we theoretically investigate the possibility that the observed 
unusual magnetism in capped gold nanoparticles is of orbital
origin. Employing semiclassical techniques, we calculate the orbital 
component to the zero-field susceptibility of individual as well as 
ensembles of metallic nanoparticles.
While the result for the orbital response of individual nanoparticles 
can exceed by orders of magnitude the bulk Landau susceptibility in 
absolute value, and can be either diamagnetic or paramagnetic depending 
on nanoparticle size, we show that the magnetic susceptibility of a
noninteracting ensemble of nanoparticles with a smooth size distribution 
is always paramagnetic at low magnetic fields. 
In particular, we predict that the zero-field susceptibility follows a 
Curie-type law for small nanoparticle sizes and/or low temperatures. 
The calculated field-dependent magnetization of an ensemble of diluted 
nanoparticles is shown to be in good agreement with existing experiments 
that yielded a large paramagnetic response. The width of the size 
distribution of the nanoparticles is identified as a key element for 
the quantitative determination of the orbital response. 
\end{abstract}

\maketitle

\section{Introduction}
Due to their small size, metallic nanoparticles show spectacular quantum 
effects that are absent in the bulk. Most of these effects stem from the 
confinement of the electronic eigenstates, which is important because of 
the relatively large surface-to-volume ratio in particles with nanometric 
sizes~\cite{kubo62_JPSJ, halpe86_RMP}. 
The most striking evidence of the quantization of the electronic states in
metallic nanoparticles is the electronic shell structure, first observed 
by Knight \textit{et al.}\ in 1984~\cite{knigh84_PRL}. 
The resulting size effects show up in many of the physical properties of 
metallic clusters, e.g., in their abundance spectra, static dipole 
polarizabilities, ionization potentials, and optical properties~\cite{heer93_RMP, brack93_RMP}. 

An aspect that attracted considerable attention over the last two decades 
is the very unusual magnetic behavior of gold nanoparticles. 
Indeed, while bulk gold is diamagnetic, several experiments have 
shown that ensembles of gold nanoparticles capped with organic ligands 
can present a ferromagneticlike behavior of the magnetization, up to 
room temperature or above \cite{cresp04_PRL, cresp06_PRL, dutta07_APL, donni07_AdvMater, garit08_NL, guerr08_Nanotechnology, guerr08, venta09, donni10_SM, maitr11_CPC, agrac17_ACSOmega}.
Other samples show a paramagneticlike behavior \cite{hori99_JPA, nakae00_PhysicaB, hori04_PRB, yamam04_PRL, yamam06, guerr08, guerr08_Nanotechnology, barto12_PRL, agrac17_ACSOmega}
and some others a diamagnetism which is typically stronger than in the bulk \cite{cresp04_PRL, dutta07_APL, guerr08_Nanotechnology, rhee13_PRL, hori04_PRB}. 
Since the experimentally-reported magnetic moments are in general very 
small, great attention has been paid to avoid spurious sources of 
magnetism in the measurements~\cite{cresp06_PRL, garci09_JAP}.
The reviews of Refs.\ \cite{nealo12_Nanoscale, donni2017} describe the
different magnetic properties that change from sample to sample, as well
as the underlying mechanisms which are at present a source of debate. 

Several mechanisms have been put forward to explain the intriguing 
magnetic properties of gold nanoparticles. 
It was proposed that the ferromagnetic response could result 
(i) from the formation of covalent bonds between the atoms residing at the 
surface of the nanoparticle and the ligands around it~\cite{cresp04_PRL}, 
(ii) from the surface atoms alone and the resulting Fermi-hole effect~\cite{hori04_PRB, yamam04_PRL}, or 
(iii) from giant electron orbits circling around single domains of ligands~\cite{herna06_PRL}.
Moreover, superconducting fluctuations that persist at temperatures which 
are orders of magnitude above the critical temperature were shown to 
result in a large diamagnetic response~\cite{imry15_PRB}, which is still 
one to two orders of magnitude smaller than the one reported in the 
experiments of Ref.\ \cite{rhee13_PRL}. 
These above-mentioned interpretations do not seem to explain all of the 
observed experimental features and are thus challenged in the literature~\cite{nealo12_Nanoscale, donni2017}. 
Moreover, the role of the molecules surrounding the nanoparticles in most 
experiments is not clear~\cite{nealo12_Nanoscale}, and ferromagnetism in 
bare gold nanoparticles has also been reported~\cite{li11_PRB}.

An alternative interpretation of the unusual magnetic properties of 
ensembles of gold nanoparticles, 
proposed by Gr\'eget \textit{et al.}\ \cite{grege12_CPC}, suggests that it 
arises from the \textit{orbital} component of the electron wave function.
Orbital magnetism is a purely quantum-mechanical effect, as stated by the 
Bohr-van Leeuwen theorem~\cite{bohr11_PhD, leeuw21_JP}. 
First studied by Landau in bulk electron gases
\cite{landa30_ZPhys, landau_statphys}, the corresponding susceptibility 
$\chi_\mathrm{L}$ equals a third of the Pauli paramagnetic spin 
susceptibility (with opposite sign), and, hence, is difficult to
measure. 
The effect of confining the electron system to a finite volume introduces 
a new energy scale in the problem (the mean level spacing) and leads to 
modifications of the Landau susceptibility. 
The investigation of these finite-size corrections included experiments 
on small metal clusters and different theoretical approaches \cite{ruite91_PRL, ruite93_MPLB, leuwe93_PhD, fraue98_PRB}. 

The diversity of the experimentally-observed behaviors recapitulated in Refs.\ \cite{nealo12_Nanoscale, donni2017}, 
as well as the distinct theoretical proposals, calls for a systematic study of 
the magnetic properties of gold nanoparticles. 
Toward this goal, we develop a theory to ponder the applicability of the 
orbital magnetism proposal to account for the experimental results. 
In particular, we seek to identify the relevant parameters of the problem, 
focusing on the temperature and size dependences of the magnetization and 
establishing in which cases a comparison with the experimental data can be 
attempted. 

Our study of orbital magnetism in metallic nanoparticles builds on previous 
works done in the mesoscopic regime of systems small enough and/or 
sufficiently cooled down to exhibit the effects of quantum coherence. 
Orbital magnetism has been experimentally and theoretically studied in this 
regime for the cases of singly- and multiply-connected geometries. 
In the latter case, when a magnetic flux pierces a 
metallic \cite{levy90_PRL, chand91_PRL} or 
semiconducting \cite{maill93_PRL} ring, the orbital response translates into 
a dissipationless persistent current~\cite{butti83_PLA}. When the unavoidable 
disorder present in these systems becomes weak enough to result in an elastic 
mean free path of the order of the sample size, the transition from a 
diffusive to a ballistic regime is achieved. 
The sustained theoretical interest in the problem of persistent currents 
during the 1990's clarified the role of disorder, electron-electron 
interactions, and the consequences of a finite number of electrons determining 
the thermodynamic functions. 
The use of the canonical ensemble appeared as 
unavoidable \cite{bouch89_JP, imry91} and a proper treatment of 
electron-electron interactions leads to an orbital response of the same 
order of magnitude as that of noninteracting systems, in both the 
diffusive \cite{schmi91_PRL, oppen91_PRL, altsh91_PRL} and the ballistic 
cases \cite{mulle93_EPL}. 
Later experiments \cite{blesz09_Science}, using a nanomechanical detection 
of persistent currents in normal-metal rings, have validated the results of 
such mean-field theories.

In the case of singly-connected geometries, the magnetic susceptibility of 
an ensemble of two-dimensional quantum dots has been experimentally 
\cite{levy93_PhysicaB} and theoretically 
\cite{oppen94_PRB, ullmo95_PRL, richt96_PhysRep} studied. 
In the ballistic regime, a semiclassical approach made it possible to obtain the 
orbital response from the magnetic field dependence of the density of states 
induced by the accumulated flux of the periodic classical trajectories. 
Interesting differences were predicted according to the chaotic or 
integrable nature of the two-dimensional underlying classical dynamics 
determined by the shape of the quantum dot boundaries. 
The orbital contribution to the magnetic susceptibility in an integrable dot 
can be diamagnetic or paramagnetic and with typical values which are orders 
of magnitude larger than the two-dimensional Landau susceptibility 
\cite{ullmo95_PRL}. Chaotic dynamics results in somehow smaller values of the 
susceptibility~\cite{richt98_EPL}.
When moving from a single quantum dot to an ensemble of dots, the average 
magnetic susceptibility was shown to be paramagnetic and smaller than the 
typical values of the individual case but still much larger than the bulk 
value \cite{richt96_PhysRep}. 
Similarly to the case of persistent currents, the inclusion of weak disorder 
\cite{richt96_PRB, richt96_JMP} or electron-electron interactions 
\cite{ullmo98_PRL} did not considerably alter the clean, noninteracting results.
 
Based on analytical semiclassical methods, together with numerical calculations, 
the mesoscopic approach presented in this paper allows us to show that the 
orbital response of an individual nanoparticle can be exceedingly large as 
compared to the bulk and either diamagnetic or paramagnetic depending on 
its size and/or Fermi level. 
In contrast, the orbital susceptibility of a statistically-distributed 
(in size) ensemble of nanoparticles is always paramagnetic at low magnetic 
fields in the absence of interactions between the nanoparticles, provided the 
size distribution is smooth and not too narrow. 
In particular, we predict that the zero-field susceptibility follows a 
Curie-type law for small nanoparticle sizes and/or low temperature. 
We further calculate the field-dependent magnetization of individual as well 
as ensembles of nanoparticles and show that the latter results are in good 
agreement with existing experiments which measured a large paramagnetic response. 

The paper is organized as follows: Section \ref{sec:model} is devoted to the
presentation of our model.
In Sec.\ \ref{sec:formalism}, we recall the semiclassical thermodynamic 
formalism that we use to evaluate the grand-canonical component of the magnetic 
response of individual nanoparticles (Sec.~\ref{sec:1NP}) and of ensembles of 
noninteracting nanoparticles with a size distribution (Sec.\
\ref{sec:manyNPs}). 
Section~\ref{sec:ind} deals with 
the magnetic response of individual nanoparticles when canonical corrections 
are taken into account. 
In Sec.\ \ref{sec:discussion}, we discuss the 
relevance of our theoretical work toward the understanding of existing 
experiments. 
We finally conclude in Sec.\ \ref{sec:ccl}. The appendixes present some details 
of our quantum (Appendix \ref{app:quantum}) and semiclassical calculations 
(Appendixes \ref{app:sc} and \ref{app:Curie}) and the basis of a possible 
extension of our model taking into account the long-ranged dipolar interaction 
between the nanoparticles of the ensemble (Appendix \ref{app:int}).

\section{Nanoparticle modeling}
\label{sec:model}
The variety of results obtained by previous works in the rich problem at hand 
arise from the multiplicity of experimental conditions and the wide window over 
which crucial physical parameters can be varied. 
In turn, the difficulties of the theoretical descriptions are a consequence of 
the previous diversity of setups and the 
necessary simplifying hypotheses to render the problem tractable. 
We start this section by clarifying the working assumptions of our 
theoretical approach, while identifying the key physical 
parameters and their range of variation. 

We assume spherical nanoparticles with radius $a$ between a few nanometers and 
a few tens of nanometers. The not too small sizes to be considered permit us to 
ignore the detailed geometrical shape of the cluster \cite{brack93_RMP} and 
allow us to use a semiclassical description \cite{gutzwiller, brack}, since for 
metallic nanoparticles we have $k_\mathrm{F}a\gg1$ (with $k_\mathrm{F}$ the 
Fermi wave vector). We choose to work with gold nanoparticles, since this is 
the case most thoroughly studied in the literature. However, a large part of 
our results are generic for any noble metal. 

The effect of the ionic background is taken into account through the use of 
the jellium approximation~\cite{brack93_RMP}. 
In addition, we treat the electron-electron interactions at mean-field level. 
The resulting self-consistent potential is approximated by a spherical well
with hard walls, thus neglecting the spill-out of electrons outside of the 
nanoparticle and the smoothness of the confining potential. 
The nanoparticles are then assumed to be large enough 
to ignore the effect of electronic correlations (which were shown to weakly 
contribute to the orbital response of disordered \cite{schmi91_PRL} and 
ballisitic \cite{ullmo98_PRL} samples) and, at the same time, smaller than 
the elastic mean-free path, such that disorder effects can be 
disregarded. 

By only describing a spin-degenerate $s$ band, we ignore the specificities 
of the electronic structure of noble metals, as well as the spin-orbit coupling. 
The calculated band structure of bulk gold indicates that the valence electrons 
can indeed be approximately treated as $s$ electrons with a parabolic
dispersion~\cite{range12_PRB} and an associated effective mass which is close 
to the bare electron mass. 
Moreover, the spin-orbit coupling has been shown not to qualitatively affect 
the magnetic response of arrays of metallic rings~\cite{mathu91_PRB}, and this 
is why we neglect such a coupling. 

The Zeeman spin splitting under a magnetic field results, in the metallic case, 
in the Pauli susceptibility $\chi_\mathrm{P}=-3\chi_\mathrm{L}$. 
Since the mesoscopic corrections to the bulk result have been shown to be 
negligible \cite{oppen94_PRB}, and since the observed effects on the zero-field 
susceptibility are typically much larger than $|\chi_\mathrm{L}|$, we do not 
consider in this work the spin effects beyond the trivial degeneracy factor. 

The ligands surrounding the nanoparticles are assumed not to play a role for 
the orbital magnetic response. 
Such a hypothesis has been challenged under the effect of particular 
ligands \cite{cresp04_PRL, yamam04_PRL, barto12_PRL, guerr08_Nanotechnology}, 
but it is generally accepted for a whole class of other protective agents 
\cite{nealo12_Nanoscale, donni2017, hori04_PRB}.

The experiments are typically performed with macroscopic samples exhibiting a 
statistical dispersion of the radius $a$ of the individual nanoparticles. 
The probability density $\mathcal{P}(a)$ characterizing such a distribution is 
a crucial element in determining the magnetic response of an ensemble of 
metallic nanoparticles. 
Often, a Gaussian probability distribution can be a good approximation to 
the experimentally observed size distribution
\cite{donni07_AdvMater, grege12_CPC, grege12_thesis}.
However, other distributions, like bimodal \cite{cresp04_PRL} or log-normal 
\cite{yamam04_PRL, dutta07_APL, barto12_PRL}, can be obtained, depending on 
the fabrication procedure. 
In addition, shell effects result in selective abundance spectra 
\cite{knigh84_PRL, heer93_RMP} and might thus lead to sharp singly- or 
multiply-peaked size distributions.

The nanoparticle concentration, and the related interparticle distance, 
is one of the important parameters of the problem. 
We will consider the case of diluted samples, where the interparticle 
interaction can be neglected. 
Temperature is another important parameter, that in the experiments is usually 
varied from cryogenic to room temperature, and we will explore 
the temperature dependence of the magnetic response in this large span in 
order to make the connection with the experimental work. 
In addition, diverse average nanoparticle sizes and size dispersions are 
typically encountered in experiments, and we show that these two parameters 
are crucial for quantitatively interpreting the experimental data which 
present a large paramagnetic response. 

Now that the assumptions used in this work have been stated and justified, 
we proceed with the presentation of our model and its Hamiltonian. 
Each spherical nanoparticle contains $N$ valence electrons with charge $-e<0$ 
and effective mass $m$. 
The nanoparticles are subject to an external, static, and homogeneous magnetic 
induction $\mathbf{B}=\nabla\times\mathbf{A}$, with $\mathbf{A}$ the associated 
vector potential. 
Within the jellium approximation~\cite{brack93_RMP}, the Hamiltonian for the 
valence electrons in an individual nanoparticle (located at the coordinate 
origin) reads in cgs units 
\begin{align}
\label{eq:H}
\mathcal{H}=&\;\sum_{i=1}^{N}
\left\{\frac{1}{2m}\left[\mathbf{p}_i+\frac{e}{c}\mathbf{A} (\mathbf{r}_i)\right]^2
+U(r_i)\right\}
\nonumber\\
&+\frac 12\sum_{\substack{i,j=1\\(i\neq j)}}^{N}V(\mathbf{r}_i, \mathbf{r}_j).
\end{align}
Here, $c$ is the speed of light, while $\mathbf{r}_i$ and $\mathbf{p}_i$ 
are the position and momentum of the 
$i$th electron, respectively. 
In Eq.\ \eqref{eq:H}, $U$ denotes the spherically-symmetric single-particle
confinement, which, for a nanoparticle in vacuum, reads as
\begin{equation}
\label{eq:U}
U(r)=\frac{Ne^2}{2a^3}\left(r^2-3a^2\right)\Theta(a-r)-\frac{Ne^2}{r}
\Theta(r-a), 
\end{equation}
i.e., it is harmonic inside the nanoparticle and Coulombic outside. 
In Eq.\ \eqref{eq:U}, $\Theta(z)$ denotes the Heaviside step function. 
Finally, $V$ represents in Eq.\ \eqref{eq:H} the Coulomb interaction amongst
electrons in the nanoparticle.
In the symmetric gauge where
$\mathbf{A}(\mathbf{r})=\frac 12 \mathbf{B}\times\mathbf{r}$, and choosing 
the $z$ axis of the coordinate system in the direction of $\mathbf{B}$, 
the Hamiltonian \eqref{eq:H} can be rewritten in the form
\begin{align}
\label{eq:H_sym}
\mathcal{H}=&\,\sum_{i=1}^N
\left[\frac{p_i^2}{2m}+U(r_i)+
\frac{\omega_\mathrm{c}}{2}l_{z,i}+\frac{m\omega_\mathrm{c}^2}{8}
\left(x_i^2+y_i^2\right)\right]
\nonumber\\
&+\frac 12\sum_{\substack{i,j=1\\(i\neq j)}}^{N}V(\mathbf{r}_i, \mathbf{r}_j), 
\end{align}
where $\omega_\mathrm{c}=eB/mc$ is the cyclotron frequency,
$\mathbf{B}=B\,\hat{\mathbf{z}}$, and $l_z$ denotes the $z$ component of the 
angular momentum.

In the sequel of the paper, we treat the electron-electron 
interactions appearing in the Hamiltonian \eqref{eq:H_sym} within
the mean-field approximation.
Density functional theory calculations~\cite{weick05_PRB, weick06_PRB} indicate 
that, in the absence of a magnetic field, the self-consistent mean-field 
potential can be approximated by 
$V_\mathrm{mf}(r)=V_0\Theta(r-a)$ where 
$V_0=E_\mathrm{F}+W$, with $E_\mathrm{F}$ and $W$ the Fermi energy and the 
work function of the considered nanoparticle, respectively.
One expects that the spherical well shape of the mean-field potential remains 
a good approximation in the presence of a magnetic field, 
provided that $\hbar\omega_\mathrm{c}$, the energy scale set by the magnetic
field, is the smallest one of the problem (for a normal metal, 
$\hbar\omega_\mathrm{c}=0.012B\,\mu\mathrm{eV/G}$)~\cite{tanak96_PRB, weick11_PRB}.
Moreover, as the magnetization is a property of the many-body ground state, 
it involves one-body states up to the vicinity of the Fermi 
level~\cite{footnote:T_F}. 
Thus, states that are higher in energy do not contribute to the magnetization. 
We can then safely assume that the height of the mean-field potential 
$V_0\to\infty$. Within these approximations, we are left with the 
effective mean-field Hamiltonian
\begin{equation}
\label{eq:H_mf}
\mathcal{H}_\mathrm{mf}=\sum_{i=1}^N
\left[\frac{p_i^2}{2m}+V_\mathrm{mf}(r_i)+
\frac{\omega_\mathrm{c}}{2}l_{z,i}+\frac{m\omega_\mathrm{c}^2}{8}
\left(x_i^2+y_i^2\right)\right]
\end{equation}
corresponding to $N$ independent electrons in a spherical billiard
threaded by a static magnetic induction in the $z$ direction. 

It is important to realize that any realistic magnetic fields that are 
experimentally available are such that the classical trajectories of the
electrons in the spherical billiard are very close to straight lines on 
the scale of the nanoparticle diameter. 
In other words, the corresponding cyclotron radius 
$R_\mathrm{c}=v_\mathrm{F}/\omega_\mathrm{c}$ ($v_\mathrm{F}$ is the
Fermi velocity) is much larger than the size of the nanoparticles we 
consider \cite{footnote:Au}.

The cylindrical symmetry of the magnetic-field dependent Hamiltonian 
\eqref{eq:H_mf} greatly facilitates its quantum-mechanical resolution. 
Furthermore, if we are only interested in the weak-field magnetic response, 
a perturbative approach can be implemented. 
Such a scheme has been successfully used in order to explain the magnetic 
response of very small metal clusters~\cite{ruite91_PRL, ruite93_MPLB, leuwe93_PhD}. 
In our case, it is important to develop simpler approaches than the quantum 
calculation, toward treating larger cluster sizes, efficiently incorporating 
the restriction of a fixed number of electrons within the nanoparticles, and 
calculating the thermodynamic functions at finite temperature. 
All of these important features of the problem at hand are readily incorporated 
within the semiclassical thermodynamic formalism presented in the next section.

\section{Semiclassical thermodynamic formalism for noninteracting nanoparticles}
\label{sec:formalism}
Here, we briefly recall the semiclassical formalism for evaluating the orbital 
susceptibility of finite-size ballistic systems (for a review, see Ref.\
\cite{richt96_PhysRep}). 
The semiclassical approach relies on the expansion of the density of states of 
the system to lowest order terms in (reduced) Planck's constant $\hbar$, which 
is a good approximation when $\hbar$ is much smaller than the action 
corresponding to the underlying classical trajectories \cite{gutzwiller, brack}. 
Such a condition is fulfilled since $k_\mathrm{F}a \gg 1$ for the nanoparticle 
sizes we consider \cite{footnote:Au}. 

For an individual nanoparticle with a \textit{fixed} number of electrons $N$ 
and at a temperature $T$, the field-dependent magnetic moment $\mathcal{M}$ and 
the zero-field susceptibility $\chi$ are given by the change of the free energy 
$F(N, T, H)$ with respect to the magnetic field $H=B-4\pi M$ ($M=\mathcal{M}/\mathcal{V}$ is the nanoparticle magnetization, with $\mathcal{V}=4\pi a^3/3$ its volume)
as
\begin{equation}
\label{eq:M_def}
\mathcal{M}=-\frac{\partial F}{\partial H}
\end{equation}
and
\begin{equation}
\label{eq:definitions}
\chi=-\frac{1}{\mathcal{V}}\left.\frac{\partial^2F}{\partial H^2}\right|_{H=0}, 
\end{equation}
respectively \cite{footnote:chi_cgs}. The use of the canonical ensemble is 
needed in order to ensure a constant number of conduction electrons in each 
nanoparticle and turns out to be crucial to obtain nonvanishing quantities 
once an ensemble average is performed~\cite{bouch89_JP, imry91, schmi91_PRL, oppen91_PRL, altsh91_PRL}. 
It is however possible and technically easier to work within the grand 
canonical ensemble with fixed chemical potential $\mu$, where the 
thermodynamic potential takes the form 
\begin{equation}
\label{eq:Omega}
\Omega(\mu, T, H)=-
k_\mathrm{B}T\int_0^\infty\mathrm{d}E\,\rho(E, H)
\ln{\left(1+\mathrm{e}^{\beta(\mu-E)}\right)}, 
\end{equation}
with $\beta=1/k_\mathrm{B}T$ the inverse temperature. The crucial quantity
entering the expression of the grand canonical potential \eqref{eq:Omega} is
the field-dependent single-particle density of states $\rho(E, H)$, which, in
a semiclassical expansion 
\cite{gutzwiller, brack, gutzw70_JMP, gutzw71_JMP, berry76_PRSL, berry77_JPA}, 
is decomposed into a mean and an oscillating (in energy) part,
$\rho(E, H)=\rho^0(E)+\rho^\mathrm{osc}(E, H)$. 

For temperatures such that $k_\mathrm{B}T$ is larger than the typical level 
spacing, $\rho^\mathrm{osc}$ can be considered as a continuous function of $E$, 
and the free energy 
\begin{equation}
\label{eq:F_Legendre}
F(N, T, H)=\Omega(\mu, T, H)+\mu N
\end{equation}
admits in the semiclassical limit the decomposition
\cite{imry91, schmi91_PRL, ullmo95_PRL, richt96_PhysRep}
\begin{equation}
\label{eq:F}
F(N)\simeq F^0+\Delta F^{(1)}+\Delta F^{(2)},
\end{equation}
where 
\begin{equation}
\label{eq:F0}
F^0=\Omega^0(\mu^0)+\mu^0 N,
\end{equation}
\begin{equation}
\label{eq:F1}
\Delta F^{(1)}=\Omega^\mathrm{osc}(\mu^0),
\end{equation}
and
\begin{equation}
\label{eq:F2}
\Delta
F^{(2)}=\frac{1}{2\rho^0(\mu^0)}
\left[\int_0^\infty\mathrm{d}E\,\rho^\mathrm{osc}(E, H)f(E)\right]^2.
\end{equation}
In Eqs.\ \eqref{eq:F0} and \eqref{eq:F1}, $\Omega^0$ and $\Omega^\mathrm{osc}$ 
are defined by using $\rho^0$ and $\rho^\mathrm{osc}$ instead of $\rho$ in 
Eq.\ \eqref{eq:Omega}, while the mean chemical potential $\mu^0$ is determined 
in such a way that the total number of electrons is
$N=\int_0^\infty\mathrm{d}E\,\rho^0(E)f(E)$, with 
$f(E)=\{\exp{(\beta[E-\mu^0])}+1\}^{-1}$ the Fermi-Dirac distribution. 
The decomposition \eqref{eq:F} results from a second-order expansion of 
Eq.\ \eqref{eq:F_Legendre} in $\mu-\mu^0$. 
In order to simplify the notation, we have only indicated the $N$ 
dependence of $F$ and the $\mu$ dependence of $\Omega$, leaving implicit 
the $T$ and $H$ dependences of both thermodynamic functions. 

Approximating the typical level spacing by the inverse of the average density 
of states 
\begin{equation}
\label{eq:rho_0}
\rho^0(E)=\frac{2\sqrt{E}}{3\pi E_0^{3/2}}
\end{equation}
taken at the Fermi energy, the condition for the previous approach to be valid 
is $(T/T_\mathrm{F})(k_\mathrm{F}a)^3\gtrsim1$, with $T_\mathrm{F}$ the Fermi 
temperature.
In Eq.~\eqref{eq:rho_0}, we defined the energy scale $E_0=\hbar^2/2ma^2$, and
a multiplicative factor of $2$ takes into account the electronic spin degeneracy.

Since, to leading order in $\hbar$, the average density of states 
\eqref{eq:rho_0} corresponds to the phase-space volume \cite{gutzwiller, brack}, 
it does not depend on the magnetic field, in agreement with the Bohr-van Leeuwen
theorem \cite{bohr11_PhD, leeuw21_JP}. 
Therefore, $F^0$ as given in Eq.~\eqref{eq:F0} does not contribute to the
magnetization at this level of approximation. 
However, higher-order terms in the $\hbar$ expansion of $\rho^0$ are 
field dependent and give rise to the three-dimensional diamagnetic Landau 
susceptibility $\chi_\mathrm{L}=-e^2k_\mathrm{F}/12\pi^2mc^2$, as can be 
shown even for constrained geometries \cite{richt96_PhysRep}
($\chi_\mathrm{L}=-2.9\times10^{-7}$ for gold).
Equation \eqref{eq:F1} yields a field-dependent term in the expansion
\eqref{eq:F} resulting in the magnetic susceptibility $\chi^{(1)}$ that would 
be obtained in the grand-canonical ensemble if the chemical potential were 
$\mu^0$. 
Equation \eqref{eq:F2} represents the ``canonical" correction to the free 
energy and leads to an additional contribution $\chi^{(2)}$ to the magnetic 
susceptibility.

\begin{figure}[tb]
\includegraphics[width=\linewidth]{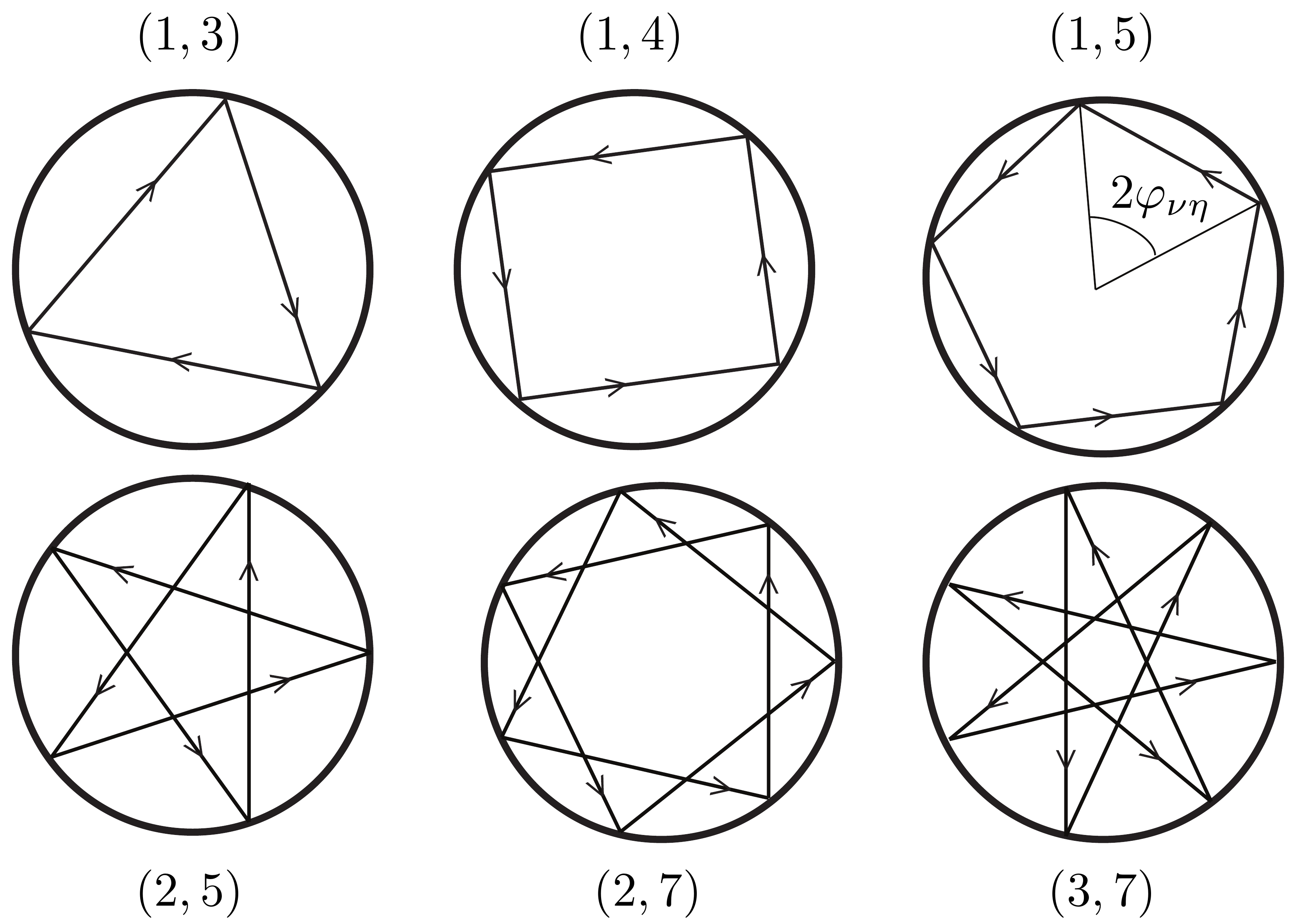}
\caption{\label{fig:trajectories}%
Example of families of classical periodic orbits on an equatorial plane of the 
sphere labeled by the topological indexes $(\nu, \eta)$, 
with $\nu$ the winding number and $\eta$ the number of bounces.}
\end{figure}

The oscillating part of the density of states corresponding to the spectrum of 
the mean-field Hamiltonian \eqref{eq:H_mf}, to first nonvanishing order in the 
magnetic field-dependent ratio $a/R_\mathrm{c}\ll1$, reads \cite{tanak96_PRB}
\begin{align}
\label{eq:rho_osc}
\rho^\mathrm{osc}(E, H)=\,&\frac{4}{E_0}\sqrt{\frac{ka}{\pi}}
\sum_{\nu=1}^\infty\sum_{\eta=2\nu+1}^\infty
\frac{(-1)^\nu
\cos{\varphi_{\nu\eta}}
\sin^{3/2}{\varphi_{\nu\eta}}
}{\sqrt{\eta}}
\nonumber\\
&\times\cos{\big(\theta_{\nu\eta}(k)\big)}
j_0\left(2\pi\phi_{\nu\eta}/\phi_0\right).
\end{align}
Here, $k=\sqrt{2mE}/\hbar$ and $j_0(z)=\sin{z}/z$ is the zeroth order spherical 
Bessel function of the first kind. 
The topological indexes $(\nu, \eta)$ label the families of classical periodic 
orbits lying on the equatorial plane of the sphere, with $\nu$ the number of 
turns around the center (i.e., the winding number) and $\eta$ the number of
specular reflections at the boundary (i.e., the number of bounces)  
\cite{footnote:DOS}. 
The quantity $\varphi_{\nu\eta}=\pi\nu/\eta$ corresponds to half the angle 
spanned between two consecutive bounces (see Fig.\ \ref{fig:trajectories}).
The length of the trajectory $(\nu, \eta)$ is given by 
$L_{\nu\eta}=2\eta a\sin{\varphi_{\nu\eta}}$. 
We further defined in Eq.\ \eqref{eq:rho_osc} the $k$-dependent phase 
$\theta_{\nu\eta}(k)=kL_{\nu\eta}+{\pi}/{4}-{3\eta\pi}/{2}$, the flux 
$\phi_{\nu\eta}=H\mathcal{A}_{\nu\eta}$ enclosed by the orbit $(\nu,\eta)$ 
covering the area 
$\mathcal{A}_{\nu\eta}=\frac 12\eta a^2\sin{(2\varphi_{\nu\eta})}$, as well as 
the flux quantum $\phi_0=hc/e$. 
Note that for the small induced fields that we encounter, $B\approx H$.

\begin{figure*}[tbh]
\includegraphics[width=\linewidth]{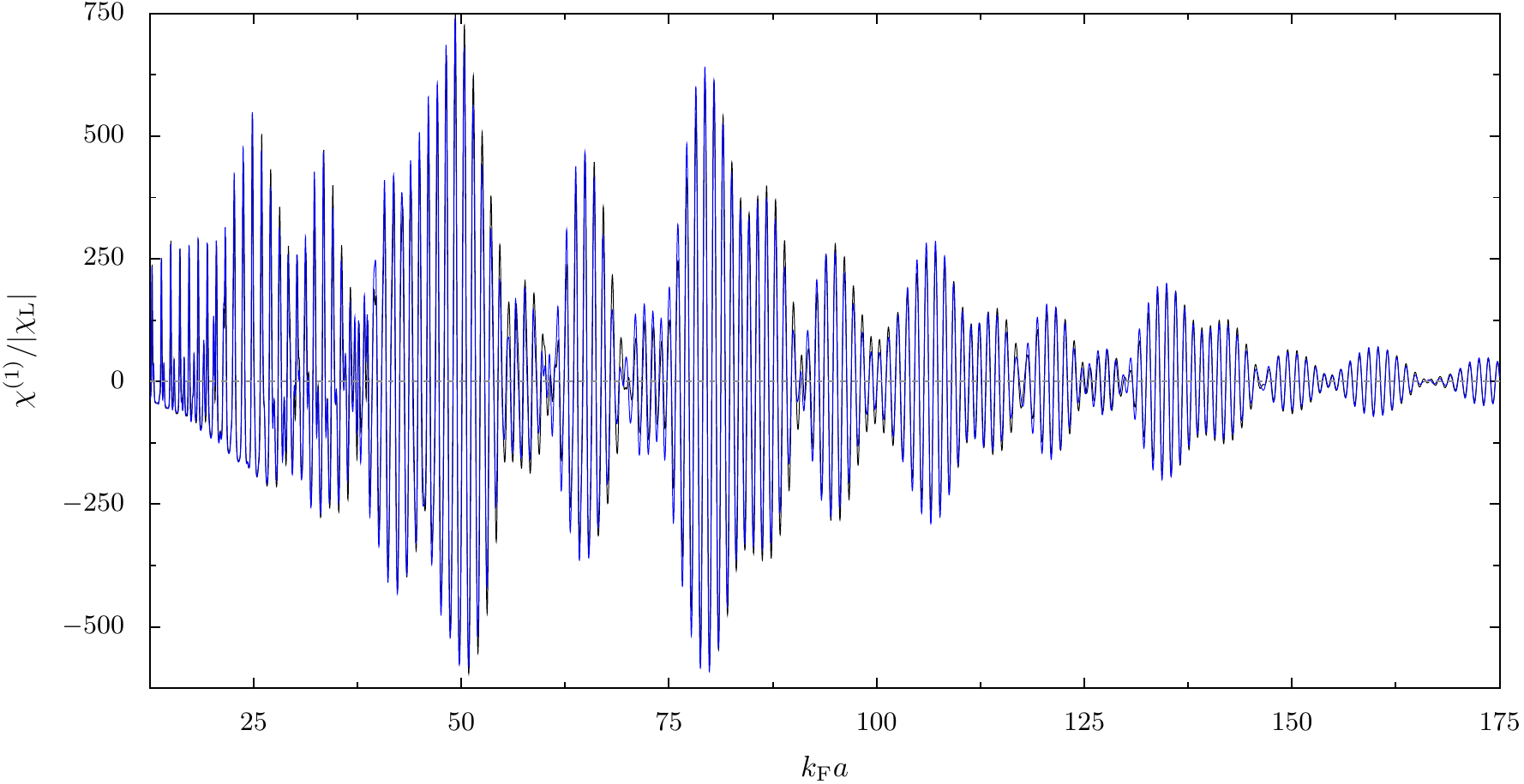}
\caption{\label{fig:chi_1NP}%
Grand-canonical zero-field susceptibility $\chi^{(1)}$, in units of the absolute 
value of the Landau susceptibility $\chi_\mathrm{L}$, as a function of the 
radius $a$ (scaled with the Fermi wave vector $k_\mathrm{F}$).
Blue line: semiclassical result from Eq.\ \eqref{eq:chi_1}. Black line: 
quantum-mechanical result from Eq.\ \eqref{eq:chi_1_QM}.
In the figure, room temperature ($T/T_\mathrm{F}=5\times10^{-3}$) is chosen 
and $\chi=0$ is indicated by the dashed gray line.}
\end{figure*}

To leading order in $k_\mathrm{F}a\gg1$, the use of 
Eq.\ \eqref{eq:rho_osc} in Eq.\ \eqref{eq:F1} yields 
\begin{align}
\label{eq:F1_result}
\Delta F^{(1)}&=
4E_\mathrm{F}\sqrt{\frac{k_\mathrm{F}a}{\pi}}
\sum_{\nu=1}^\infty\sum_{\eta=2\nu+1}^\infty
\frac{(-1)^\nu\cos{\varphi_{\nu\eta}}}{\eta^{5/2}\sqrt{\sin{\varphi_{\nu\eta}}}}
\nonumber\\
&\times
R(L_{\nu\eta}/L_T) 
\cos{\big(\theta_{\nu\eta}(k_\mathrm{F})\big)}
j_0\left(2\pi\phi_{\nu\eta}/\phi_0\right).
\end{align}
In the above expression, the thermal factor
\begin{equation}
\label{eq:R_T}
R(L/L_T)=\frac{L/L_T}{\sinh{(L/L_T)}}
\end{equation}
arises from the energy integration, and exponentially suppresses the 
zero-temperature contribution of each family of trajectories with length 
$L$ according to the ratio $L/L_T$, where 
$L_T=\hbar v_\mathrm{F}/\pi k_\mathrm{B}T$ is the thermal length.
In a similar fashion, the energy integral of Eq.\ \eqref{eq:F2} leads to 
the second-order correction
\begin{align}
\label{eq:F2_result}
\Delta F^{(2)}=&\,
12E_\mathrm{F}
\Bigg[\sum_{\nu=1}^\infty\sum_{\eta=2\nu+1}^\infty
\frac{(-1)^\nu
\cos{\varphi_{\nu\eta}}
\sqrt{\sin{\varphi_{\nu\eta}}}
}{\eta^{3/2}}
\nonumber\\
&\times
R(L_{\nu\eta}/L_T)
\sin{\big(\theta_{\nu\eta}(k_\mathrm{F})\big)}
j_0\left(2\pi\phi_{\nu\eta}/\phi_0\right)
\Bigg]^2. 
\end{align}
In evaluating Eqs.\ \eqref{eq:F1} and \eqref{eq:F2}, we identified $\mu^0$ 
with $E_\mathrm{F}$, neglecting the temperature correction to the chemical 
potential which is of order $(T/T_\mathrm{F})^2\ll1$.

The canonical correction \eqref{eq:F2_result} to the free energy is an 
order $\sqrt{k_\mathrm{F}a}$ lower than the grand-canonical contribution 
\eqref{eq:F1_result}. 
The condition $\Delta F^{(2)}\ll |\Delta F^{(1)}|$, on which the validity of 
the decomposition \eqref{eq:F} is based, then reposes on a more stringent 
constraint than that of the semiclassical approximation ($k_\mathrm{F}a\gg1$). 
The fulfillment of the condition 
$\Delta F^{(2)} \ll |\Delta F^{(1)}|$ translates into 
$|\chi^{(2)}| \ll |\chi^{(1)}|$ for sufficiently large $k_\mathrm{F}a$, 
but the previous inequality might not hold for moderate values of 
$k_\mathrm{F}a$ (in the same way as we may have $|\Delta F^{(1)}| \ll F^0$ and 
$|\chi^{(1)}|\gg |\chi_\mathrm{L}|$).
When $|\chi^{(2)}| \ll |\chi^{(1)}|$, the orbital response of an individual 
nanoparticle is then dominated by the grand-canonical contribution. 
However, as we will see, in certain cases the latter may become 
negligible once the average over an ensemble of nanoparticles is performed. 
Thus, Eq.\ \eqref{eq:F2_result} is crucial to obtain nonvanishing quantities 
for the resulting magnetic response of an ensemble of noninteracting 
nanoparticles with an important size dispersion (see Sec.\ \ref{sec:manyNPs}).

Using the leading-in-$\hbar$, field-dependent contribution \eqref{eq:F1_result} 
to the free energy, the grand-canonical contribution to the magnetic moment 
[see Eq.\ \eqref{eq:M_def}] is given by the semiclassical expression
\begin{align}
\label{eq:M_1}
\frac{\mathcal{M}^{(1)}}{\mu_\mathrm{B}}=&\,-\frac{4}{\sqrt{\pi}}(k_\mathrm{F}a)^{5/2}
\sum_{\substack{\nu>0\\\eta>2\nu}}
\frac{(-1)^\nu
\cos^2{\varphi_{\nu\eta}}
\sqrt{\sin{\varphi_{\nu\eta}}}
}{\eta^{3/2}}
\nonumber\\
&\times
R(L_{\nu\eta}/L_T) 
\cos{\big(\theta_{\nu\eta}(k_\mathrm{F})\big)}
j'_0\left(2\pi\phi_{\nu\eta}/\phi_0\right)
\end{align}
in terms of the Bohr magneton $\mu_\mathrm{B}=e\hbar/2mc$. Here, $j_0'(z)$ 
denotes the derivative of $j_0(z)$ with respect to $z$.
The corresponding zero-field susceptibility is \cite{fraue98_PRB}
\begin{align}
\label{eq:chi_1}
\frac{\chi^{(1)}}{|\chi_\mathrm{L}|}=&\;6\sqrt{\pi}(k_\mathrm{F}a)^{3/2}
\sum_{\substack{\nu>0\\\eta>2\nu}}
\frac{(-1)^\nu
\cos^3{\varphi_{\nu\eta}}
\sin^{3/2}{\varphi_{\nu\eta}}
}{\sqrt{\eta}}
\nonumber\\
&\times
R(L_{\nu\eta}/L_T) 
\cos{\big(\theta_{\nu\eta}(k_\mathrm{F})\big)}.
\end{align}

Similarly, from Eq.\ \eqref{eq:F2_result} we obtain the semiclassical 
expressions for the canonical contribution to the magnetic moment
\begin{align}
\label{eq:M_2}
\frac{\mathcal{M}^{(2)}}{\mu_\mathrm{B}}=&\,
-24 (k_\mathrm{F}a)^2
\sum_{\substack{\nu>0\\\eta>2\nu}}\sum_{\substack{\nu'>0\\\eta'>2\nu'}}
\frac{\mathcal{F}_{\nu\eta}^{\nu'\eta'}}{\eta\cos{\varphi_{\nu\eta}}\sin{\varphi_{\nu\eta}}}
\nonumber\\
&\times
R(L_{\nu\eta}/L_T) R(L_{\nu'\eta'}/L_T) 
\nonumber\\
&\times
\sin{\big(\theta_{\nu\eta}(k_\mathrm{F})\big)}
\sin{\big(\theta_{\nu'\eta'}(k_\mathrm{F})\big)}
\nonumber\\
&\times
j'_0\left(2\pi\phi_{\nu\eta}/\phi_0\right)j_0\left(2\pi\phi_{\nu'\eta'}/\phi_0\right)
\end{align}
and the zero-field susceptibility
\begin{align}
\label{eq:chi_2}
\frac{\chi^{(2)}}{|\chi_\mathrm{L}|}=&\;
36\pi k_\mathrm{F}a
\sum_{\substack{\nu>0\\\eta>2\nu}}\sum_{\substack{\nu'>0\\\eta'>2\nu'}}
\mathcal{F}_{\nu\eta}^{\nu'\eta'}
\nonumber\\
&\times
R(L_{\nu\eta}/L_T) R(L_{\nu'\eta'}/L_T) 
\nonumber\\
&\times
\sin{\big(\theta_{\nu\eta}(k_\mathrm{F})\big)}
\sin{\big(\theta_{\nu'\eta'}(k_\mathrm{F})\big)}.
\end{align}
In Eqs.\ \eqref{eq:M_2} and \eqref{eq:chi_2} we have defined 
\begin{align}
\label{eq:f}
\mathcal{F}_{\nu\eta}^{\nu'\eta'}=&\;(-1)^{\nu+\nu'}\eta^{1/2}\eta'^{-3/2}
\cos^3{\varphi_{\nu\eta}}\cos{\varphi_{\nu'\eta'}}
\nonumber\\
&\times
\sin^{5/2}{\varphi_{\nu\eta}}\sin^{1/2}{\varphi_{\nu'\eta'}}.
\end{align}
In the following sections we will evaluate the previous semiclassical 
expressions in different parameter regimes.

\section{Grand-canonical magnetic response}
\label{sec:1NP}
The grand-canonical sums \eqref{eq:M_1} and \eqref{eq:chi_1} over the 
topological indexes can be readily evaluated numerically since the thermal 
factor \eqref{eq:R_T} acts as a cutoff for long trajectories, keeping us 
away from the typical convergence problems of semiclassical expansions. 
At the practical level, we perform the sums by only retaining trajectories 
that are shorter than $10 L_T$, and since the sum over $\eta$ converges
relatively fast (the summand decreases as $1/\eta^2$ when $\eta\gg\nu$), 
we perform it up to $\eta_\mathrm{max}=100\nu$ (for a given $\nu$). 
We have checked that including trajectories with larger $\nu$ and/or $\eta$ 
does not lead to significant changes in the final results. 
 
The zero-field susceptibility \eqref{eq:chi_1} is shown in 
Fig.\ \ref{fig:chi_1NP} as a blue solid line as a function of the size $a$ 
for a temperature $T/T_\mathrm{F}=5\times10^{-3}$ that approximately 
corresponds to room temperature \cite{footnote:Au}.
As can be seen from the figure, $\chi^{(1)}$ oscillates and changes sign as 
a function of $k_\mathrm{F}a$. 
Moreover, the magnetic susceptibility can take values that are much larger 
than the magnitude of the Landau value $|\chi_\mathrm{L}|$. 
Depending on the nanoparticle size, large paramagnetic or diamagnetic 
responses can be obtained.
The rapidly oscillating behavior of the zero-field susceptibility as a 
function of the sphere radius stems from the dependence of the density of 
states on the action of the dominant periodic orbits. A similar behavior has 
been found in two dimensions \cite{ullmo95_PRL, richt96_PhysRep}, and also 
the prefactor $(k_\mathrm{F}a)^{3/2}$ of $\chi^{(1)}$ in Eq.\ \eqref{eq:chi_1} 
is in line with the two-dimensional case.
The beating pattern present in the susceptibility $\chi^{(1)}$ is due to 
interferences between periodic trajectories of different length. 
The overall amplitude of these beatings decays for the largest sizes due to 
the thermal factor \eqref{eq:R_T} appearing in Eq.\ \eqref{eq:chi_1}, 
such that $\lim_{k_\mathrm{F}a\rightarrow\infty}\chi^{(1)}=0$. 
Within this limit, one thus recovers the Landau bulk susceptibility 
$\chi_\mathrm{L}$ for the total orbital susceptibility of the system.
 
\begin{figure}[tb]
\includegraphics[width=\linewidth]{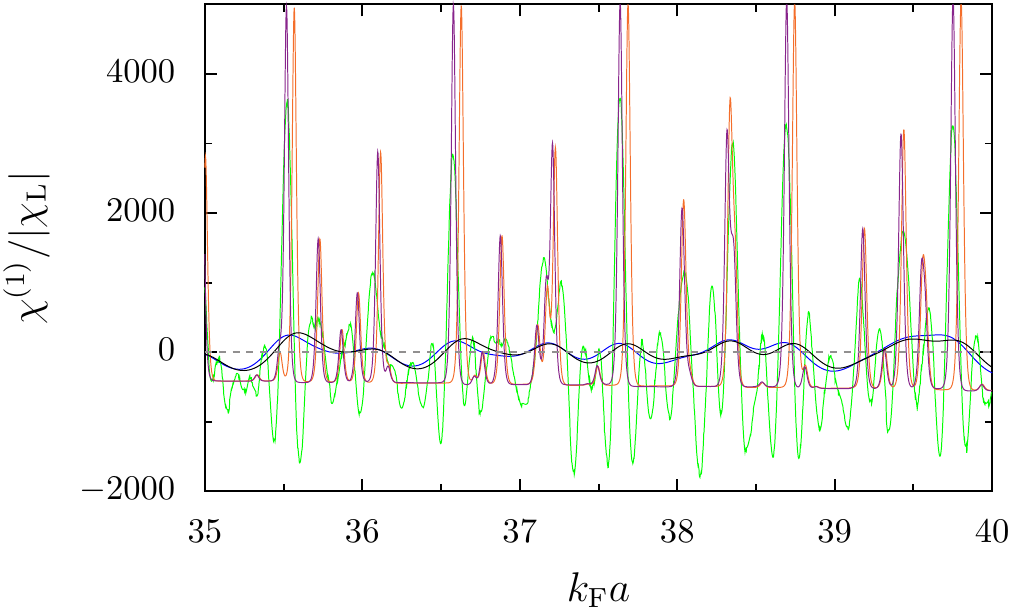}
\caption{\label{fig:chi_1NP_lowT}%
Same quantity from Eq.\ \eqref{eq:chi_1} as in Fig.\ \ref{fig:chi_1NP},
in a restricted $a$ interval for $T/T_\mathrm{F}=5\times10^{-3}$ (blue line) and
$T/T_\mathrm{F}=5\times10^{-4}$ (violet line).
The quantum-mechanical result \eqref{eq:chi_1_QM} (black line: 
$T/T_\mathrm{F}=5\times10^{-3}$; orange line: $T/T_\mathrm{F}=5\times10^{-4}$) 
and approximate semiclassical result \eqref{eq:chi_1_SC} for 
$T/T_\mathrm{F}=5\times10^{-4}$
(green line) are shown for comparison purposes.}
\end{figure}

That the result of the semiclassical sum \eqref{eq:chi_1} with the 
above-explicited approximations gives a good account of the quantum results 
can be checked in the parameter range accessible to 
both approximations (compare the blue and black lines in 
Fig.\ \ref{fig:chi_1NP}, which are almost indistinguishable on this 
large-scale figure, and the violet and orange lines in 
Fig.\ \ref{fig:chi_1NP_lowT}).
The perturbative quantum calculation (to second order in the magnetic field), 
limited to small clusters and low temperatures, results from a numerical
evaluation over the eigenstates of the unperturbed problem  
\cite{ruite91_PRL, leuwe93_PhD} (see Appendix \ref{app:quantum} for details). 
 
While the previous agreement is not surprising, given that 
Fig.\ \ref{fig:chi_1NP} presents results in the semiclassical limit 
$k_\mathrm{F}a\gg1$ for high (room) temperature, 
Fig.\ \ref{fig:chi_1NP_lowT} shows that at low temperatures 
($T/T_\mathrm{F}=5\times10^{-4}$) the semiclassical sum \eqref{eq:chi_1} 
also reproduces the quantum result \eqref{eq:chi_1_QM}. 
The paramagnetic peaks, with values that exceed the Landau susceptibility 
by orders of magnitude, are observed at the eigenenergies of the unperturbed 
system, while the negative (diamagnetic) background is given by the small 
quadratic (in magnetic field) contribution represented by the last term on the 
right-hand side of Eq.~\eqref{eq:chi_1_QM}. 
Although not visible on the scale of Fig.\ \ref{fig:chi_1NP_lowT}, the 
diamagnetic background increases with $k_\mathrm{F}a$ due to the incorporation 
of more states in the sums as the Fermi energy increases. 
The dependence of the energy levels on the applied magnetic field 
discussed in Appendix \ref{app:quantum} and shown in 
Fig.\ \ref{fig:exact_diag} allows for an understanding of the peak structure in the susceptibility that is found at low temperatures (see Fig.\ \ref{fig:chi_1NP_lowT}). The positive curvature of the individual 
levels yields the diamagnetic background that becomes stronger when more 
levels are occupied. 
The crossings of levels with different magnetic quantum number 
at zero applied field translate in a diverging negative curvature of the total 
energy and a corresponding paramagnetic peak when the chemical 
potential coincides with such a level crossing. 
Temperature smears the peaks and limits their height due to admixtures 
of contributions from neighboring levels. The rapid oscillations of the 
susceptibility found at room temperature as a function of the chemical 
potential and/or sphere radius are the remainders of that peak structure.
It is remarkable that a semiclassical expansion like that of 
Eq.~\eqref{eq:chi_1} is able to reproduce signatures characteristic of 
individual eigenenergies. We notice, however, that each energy represents 
$2(2l+1)$ degenerate unperturbed states, with $l$ the angular momentum 
quantum number, and that very long trajectories have to be included in the 
semiclassical calculation to approach the quantum result of 
Fig.\ \ref{fig:chi_1NP_lowT}. 
  
The semiclassical sum \eqref{eq:chi_1} may be challenging to implement at low 
temperature, due to the non-negligible contribution from very long trajectories 
to $\chi^{(1)}$. It is then useful to further develop the semiclassical 
expansion \eqref{eq:chi_1} by an approximate analytical calculation. 
Such a calculation, presented in Appendix \ref{app:sc}, relies on trading 
the thermal factor \eqref{eq:R_T} by a Heaviside function that limits the 
contributing trajectories to the maximal length $L_\mathrm{max}=\alpha L_T$ 
and performs the $\nu$ sum by Poisson summation rule, followed by a 
stationary-phase approximation. 
The cutoff length $L_\mathrm{max}$ is chosen as that in which the thermal 
factor \eqref{eq:R_T} presents the maximum derivative, yielding 
$\alpha\simeq1.6$. 
When the thermal factor is replaced by $\Theta(L_\mathrm{max}-L_{\nu\eta})$,
such a value of $\alpha$ yields at low temperature results for $\chi^{(1)}$ in 
excellent agreement with the original expression \eqref{eq:chi_1}.
The resulting magnetic susceptibility is then given in the limit 
$k_\mathrm{F}a\frac{T}{T_\mathrm{F}}\ll1$ (keeping $k_\mathrm{F}a\gg1$) by 
\begin{equation}
\label{eq:chi_1_SC}
 \frac{\chi^{(1)}}{|\chi_\mathrm{L}|}\simeq\frac{3}{4(k_\mathrm{F}a)^2}\sum_{\eta=3}^\infty
 \sum_{\substack{j=j_\mathrm{min}\\ (j\ \mathrm{odd})}}^{j_\mathrm{max}}
 j^3\sqrt{1-\left(\frac{j}{2k_\mathrm{F}a}\right)^2}\cos\left(\eta S_j \right), 
\end{equation}
where the phase factor $S_j$, which corresponds to the (dimensionless) radial 
action, is defined as
\begin{equation}
\label{eq:S_j}
S_j = \sqrt{(2k_\mathrm{F}a)^2-j^2}- j\arccos\left(\frac{j}{2k_\mathrm{F}a} \right) -\frac{3\pi}{2}.
\end{equation}
In Eq.\ \eqref{eq:chi_1_SC}, the summation over $j$ (which must be an odd 
integer) depends on the value of $\eta$. 
For $3\leqslant\eta\leqslant\eta_\mathrm{c}$, with 
$\eta_\mathrm{c}=\alpha L_T/2a=(\alpha/\pi) 
(k_\mathrm{F}a\frac{T}{T_\mathrm{F}})^{-1}$, we have $j_\mathrm{min}=1$ and 
$j_\mathrm{max}=\lfloor 2k_\mathrm{F}a\cos{\vartheta_\eta}\rfloor$ with 
$\vartheta_\eta=\pi/2\eta$ if $\eta$ is odd and
$j_\mathrm{min}=\lceil2k_\mathrm{F}a\sin{\vartheta_\eta}\rceil$ and  
$j_\mathrm{max}=\lfloor 2k_\mathrm{F}a\cos{\vartheta_\eta}\rfloor$ 
if $\eta$ is even.
For $\eta>\eta_\mathrm{c}$, we have $j_\mathrm{min}=\lceil 2k_\mathrm{F}a\cos{\left(\arcsin\left(\eta_\mathrm{c}/\eta\right)+\vartheta_\eta\right)}\rceil$ 
and $j_\mathrm{max}=\lfloor 2k_\mathrm{F}a\cos{\vartheta_\eta}\rfloor$. 
Here, $\lfloor x \rfloor$ and $\lceil x\rceil$ denote the floor and ceiling 
functions, respectively.

The sum \eqref{eq:chi_1_SC} is considerably simpler to implement, as compared 
with that of Eq.\ \eqref{eq:chi_1}, and gives rather accurate results for low 
temperatures and/or small nanoparticle sizes (see the green line in 
Fig.\ \ref{fig:chi_1NP_lowT}). 
For high temperatures, the sharp cutoff imposed when $L>L_\mathrm{max}$ is a 
too restrictive approximation that ignores the exponential fall off of the 
thermal factor~\eqref{eq:R_T}, and the previous agreement deteriorates. 
Nevertheless, in this regime the evaluation of Eq.~\eqref{eq:chi_1} is again
simple, since we only need to include the contribution of the shortest 
trajectories with a winding number of $\nu=1$ and the appropriate 
exponential fall off resulting from $R(L_{1\eta}/L_T)$ (results not shown). 

\begin{figure}[tb]
\includegraphics[width=\linewidth]{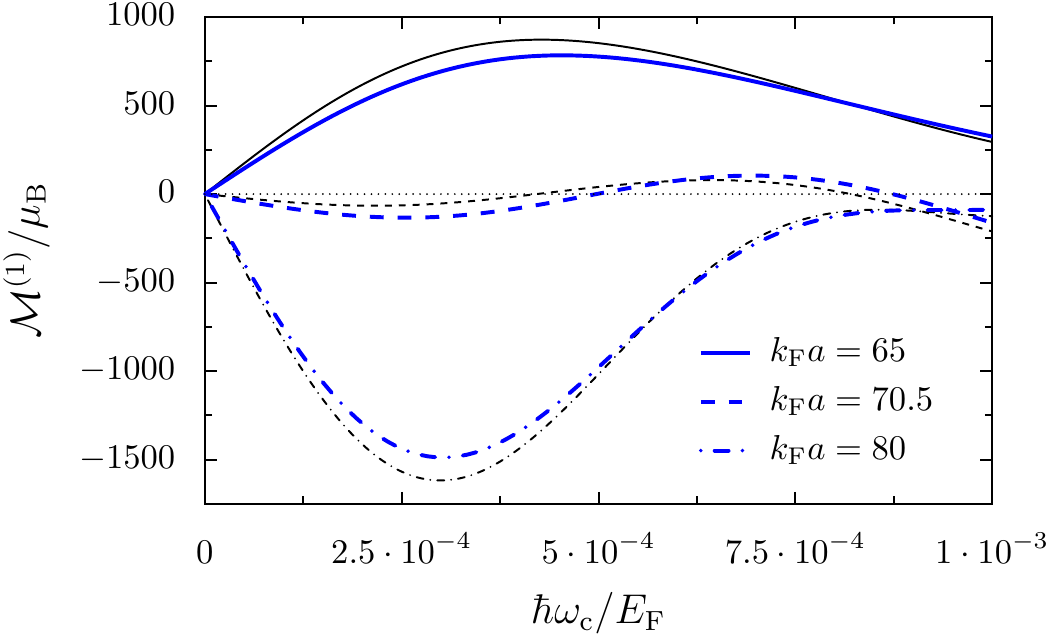}
\caption{\label{fig:M_1NP}%
Grand-canonical magnetic moment $\mathcal{M}^{(1)}$ in units of the Bohr 
magneton $\mu_\mathrm{B}$ for three different nanoparticle sizes as a function 
of the cyclotron frequency $\omega_\mathrm{c}\propto H$ (in units of 
$E_\mathrm{F}/\hbar$).
Blue lines: semiclassical result from Eq.\ \eqref{eq:M_1}. 
Black lines: perturbative quantum result from Eq.\ \eqref{eq:M_1_QM}.
In the figure, $T/T_\mathrm{F}=5\times10^{-3}$.}
\end{figure}
The grand-canonical finite-field magnetization according to the semiclassical 
expression \eqref{eq:M_1} is presented in Fig.\ \ref{fig:M_1NP}
as a function of the cyclotron frequency $\omega_\mathrm{c}\propto H$ 
(blue lines).
The range of $\hbar\omega_\mathrm{c}/E_\mathrm{F}$ corresponds to realistic
values of the magnetic field that are at present experimentally available 
(for Au, $\hbar\omega_\mathrm{c}/E_\mathrm{F}=10^{-3}$ corresponds to a field 
of the order of $H=\unit[45\times10^4]{Oe}$).
The different slopes at the origin obtained for the selected values of $a$ are 
in line with the rapid oscillations of $\chi^{(1)}$ as a function of size 
(see Fig.\ \ref{fig:chi_1NP}). 
The diamagnetic or paramagnetic character of the zero-field susceptibility 
might change at finite fields due to the possible nonmonotonic behavior of 
$\mathcal{M}^{(1)}(H)$ and its possible sign inversion for particular values 
of $k_\mathrm{F}a$ (see dashed lines in Fig.~\ref{fig:M_1NP}). 
Large values of the magnetic moment (of several hundreds of $\mu_\mathrm{B}$) 
can be attained. 
We further show in Fig.\ \ref{fig:M_1NP} by black lines the perturbative 
quantum result from Eq.\ \eqref{eq:M_1_QM}. As it is the case for the 
zero-field susceptibility shown in Figs.~\ref{fig:chi_1NP} and 
\ref{fig:chi_1NP_lowT}, the semiclassical result gives a very good 
qualitative account of the quantum one.

\section{Magnetic response of an ensemble of noninteracting nanoparticles}
\label{sec:manyNPs}
The experiments yielding unusual magnetism in gold nanoparticles are typically 
performed on ensembles of nanoparticles~\cite{nealo12_Nanoscale, donni2017}.
We thus consider in this section the orbital response of such ensembles, 
neglecting any possible interparticle interaction. 
This approximation should be valid in relatively dilute samples. 

For an ensemble of $\mathcal{N}$ nanoparticles, the expected value of the 
zero-field susceptibility is 
\begin{equation}
\label{eq:chi_ens_def}
\chi_\mathrm{ens}(\bar a, \delta a)=\overline{\chi^{(1)}}+\overline{\chi^{(2)}}, 
\end{equation}
while the root-mean-square deviation with respect to the previous value is
\begin{equation}
\label{eq:rmsd}
\chi_\mathrm{rmsd}\simeq \frac{1}{\sqrt{\mathcal{N}}}
\left[\overline{\left(\chi^{(1)}\right)^2}\right]^{1/2}.
\end{equation}
The averages indicated by a bar are taken with respect to a probability 
distribution of sizes $\mathcal{P}(a)$. 
In writing Eq.\ \eqref{eq:rmsd}, we have used the fact that the typical 
values of $\chi^{(1)}$ are much larger than those of $\chi^{(2)}$, 
which is valid for sufficiently large values of $k_\mathrm{F}a$ and 
$T/T_\mathrm{F}$. 

The magnetic response of an ensemble of nanoparticles crucially depends on 
its size distribution. 
The large diversity that can be encountered for the latter is at the origin 
of the rich range of observed physical behaviors. 
In order to provide quantitative predictions, we will focus on setups 
well described by a Gaussian probability distribution 
\begin{equation}
\label{eq:P(a)}
\mathcal{P}(a)=\frac{1}{\sqrt{2\pi}\delta a}
\exp{\left(-\frac{(a-\bar a)^2}{2 {\delta a}^2}\right)}, 
\end{equation}
characterized by the average radius $\bar{a}$ of the ensemble and its size 
dispersion $\delta a$. 

The rapidly oscillating cosine in Eq.\ \eqref{eq:chi_1} 
(see Figs.\ \ref{fig:chi_1NP} and \ref{fig:chi_1NP_lowT}) results in a 
$\overline{\chi^{(1)}}$ which decreases exponentially with 
$k_\mathrm{F}\delta a$ and is thus much smaller than 
$|\chi_\mathrm{L}|$ when the size dispersion 
$\delta a\gtrsim k_\mathrm{F}^{-1}\sim\unit[1]{\AA}$. 
In situations where the dispersion $\delta a$ is larger than $\unit[1]{\AA}$, 
as is usually the case in experiments~\cite{nealo12_Nanoscale, grege12_thesis}, 
$\overline{\chi^{(1)}}$ is therefore negligible. 
It is thus $\overline{\chi^{(2)}}$ which yields the dominant contribution to
the averaged magnetic susceptibility of the ensemble. 
Similar considerations and definitions hold for the magnetic moment per particle.
The identification of $\chi_\mathrm{ens}$ with the measure on an ensemble of 
$\mathcal{N}$ nanoparticles is statistically sound only for a 
sufficiently large $\mathcal{N}$ such that 
$\chi_\mathrm{rmsd}\ll\chi_\mathrm{ens}$. 
There are then two parameters that might result in large variations of the 
zero-field susceptibility: the size dispersion $\delta a$ and the number of 
nanoparticles~$\mathcal{N}$.

Averaging $\mathcal{M}^{(2)}$ and $\chi^{(2)}$ [cf.\ Eqs.\ \eqref{eq:M_2} 
and \eqref{eq:chi_2}] over the Gaussian distribution \eqref{eq:P(a)} 
(for $k_\mathrm{F}\delta a\gtrsim1$), we obtain 
\begin{align}
\label{eq:M_ens}
\frac{\mathcal{M}_\mathrm{ens}(\bar a, \delta a)}
{\mu_\mathrm{B}}=&\,-12(k_\mathrm{F}\bar a)^2
\sum_{\substack{\nu>0\\\eta>2\nu}}
\sum_{\substack{\nu'>0\\\eta'>2\nu'}}
\frac{\mathcal{F}_{\nu\eta}^{\nu'\eta'}}
{\eta\cos{\varphi_{\nu\eta}}\sin{\varphi_{\nu\eta}}}
\nonumber\\
&\times
R(L_{\nu\eta}/L_T) R(L_{\nu'\eta'}/L_T) 
\nonumber\\
&\times
\cos{\big(\theta_{\nu\eta}(k_\mathrm{F})-\theta_{\nu'\eta'}(k_\mathrm{F})\big)}
\nonumber\\
&\times
j'_0\left(2\pi\phi_{\nu\eta}/\phi_0\right)
j_0\left(2\pi\phi_{\nu'\eta'}/\phi_0\right)
\nonumber\\
&\times
\mathrm{e}^{-2[k_\mathrm{F}\delta a
(\eta\sin{\varphi_{\nu\eta}}-\eta'\sin{\varphi_{\nu'\eta'}})]^2}
\end{align}
and
\begin{align}
\label{eq:chi_ens}
\frac{\chi_\mathrm{ens}(\bar a, \delta a)}{|\chi_\mathrm{L}|}
=&\;18\pi k_\mathrm{F}\bar a
\sum_{\substack{\nu>0\\\eta>2\nu}}
\sum_{\substack{\nu'>0\\\eta'>2\nu'}}
\mathcal{F}_{\nu\eta}^{\nu'\eta'}
\nonumber\\
&\times
R(L_{\nu\eta}/L_T) R(L_{\nu'\eta'}/L_T) 
\nonumber\\
&\times
\cos{\big(\theta_{\nu\eta}(k_\mathrm{F})-\theta_{\nu'\eta'}(k_\mathrm{F})\big)}
\nonumber\\
&\times
\mathrm{e}^{-2[k_\mathrm{F}\delta
a(\eta\sin{\varphi_{\nu\eta}}-\eta'\sin{\varphi_{\nu'\eta'}})]^2}, 
\end{align}
respectively.
In Eqs.\ \eqref{eq:M_ens} and \eqref{eq:chi_ens}, the quantities $L_{\nu\eta}$ 
and $\theta_{\nu\eta}$ are evaluated for $a=\bar{a}$, and 
$\mathcal{F}_{\nu\eta}^{\nu'\eta'}$ is defined in Eq.\ \eqref{eq:f}.

\begin{figure}[tb]
\includegraphics[width=\linewidth]{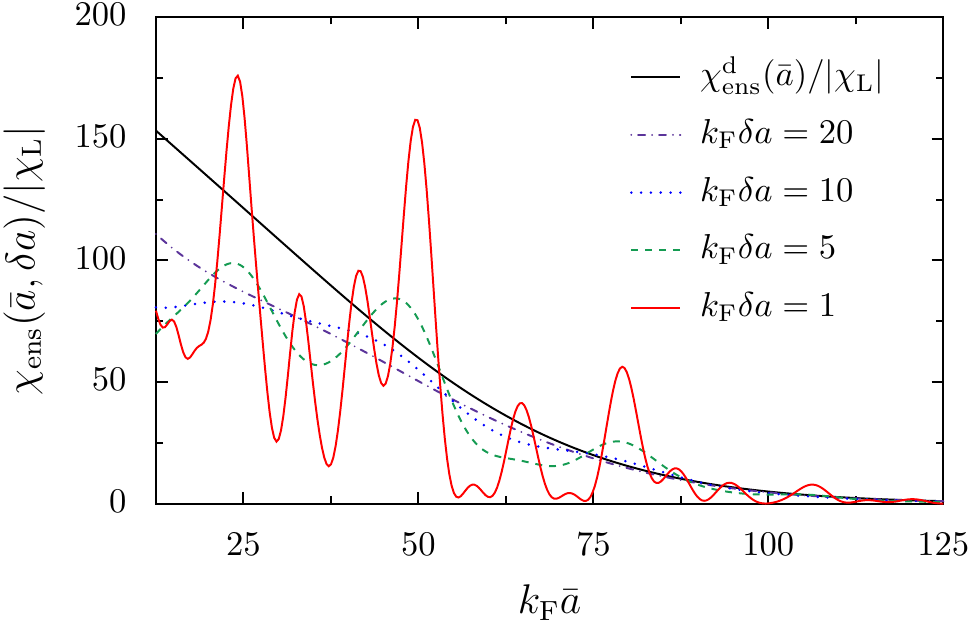}
\caption{\label{fig:chi_ens}%
Zero-field susceptibility of a nanoparticle ensemble with average radius 
$\bar{a}$ for various size dispersions $\delta a$ at 
$T/T_\mathrm{F}=5\times10^{-3}$, from Eq.\ \eqref{eq:chi_ens}. 
The diagonal contribution to the averaged susceptibility
$\chi_\mathrm{ens}^\mathrm{d}(\bar a)$ [Eq.\ \eqref{eq:chi_scaling}] is
shown for comparison as a black solid line.}
\end{figure}
Sums like \eqref{eq:M_ens} and \eqref{eq:chi_ens}, running over four topological 
indexes (corresponding to two different families of periodic orbits), are even 
more challenging to evaluate than those running over two indexes, 
as Eqs.\ \eqref{eq:M_1} and \eqref{eq:chi_1}, especially at low temperatures, 
where a large number of classical trajectories has to be considered. 
The ensemble-averaged zero-field susceptibility resulting from 
Eq.\ \eqref{eq:chi_ens} at high (room) temperature 
$T/T_\mathrm{F}=5\times10^{-3}$ is presented in Fig.\ \ref{fig:chi_ens} 
as a function of the average nanoparticle radius $\bar{a}$, for increasing 
size dispersions $\delta a$.
The orbital response of the nanoparticle ensemble at zero magnetic field 
is \textit{paramagnetic} ($\chi_\mathrm{ens}>0$) in all tested cases.
As discussed in the introduction, such is also the case in ensembles of 
quasi-two-dimensional semiconductor quantum 
dots~\cite{levy93_PhysicaB, oppen94_PRB, ullmo95_PRL, richt96_PhysRep, richt96_PRB, richt96_JMP}.
The orbital susceptibility of the ensemble $\chi_\mathrm{ens}$ can reach 
large values (in units of $|\chi_\mathrm{L}|$) for not too large mean radii, 
but it goes to zero when $k_\mathrm{F}\bar{a}\gg1$.
The monotonic decrease of $\chi_\mathrm{ens}$ with $k_\mathrm{F}\bar{a}$ 
obtained for large size dispersions ($k_\mathrm{F}\delta a\gtrsim 20$ in 
Fig.~\ref{fig:chi_ens}) evolves into an oscillating behavior for smaller 
size dispersions.

\begin{figure}[tb]
\includegraphics[width=\linewidth]{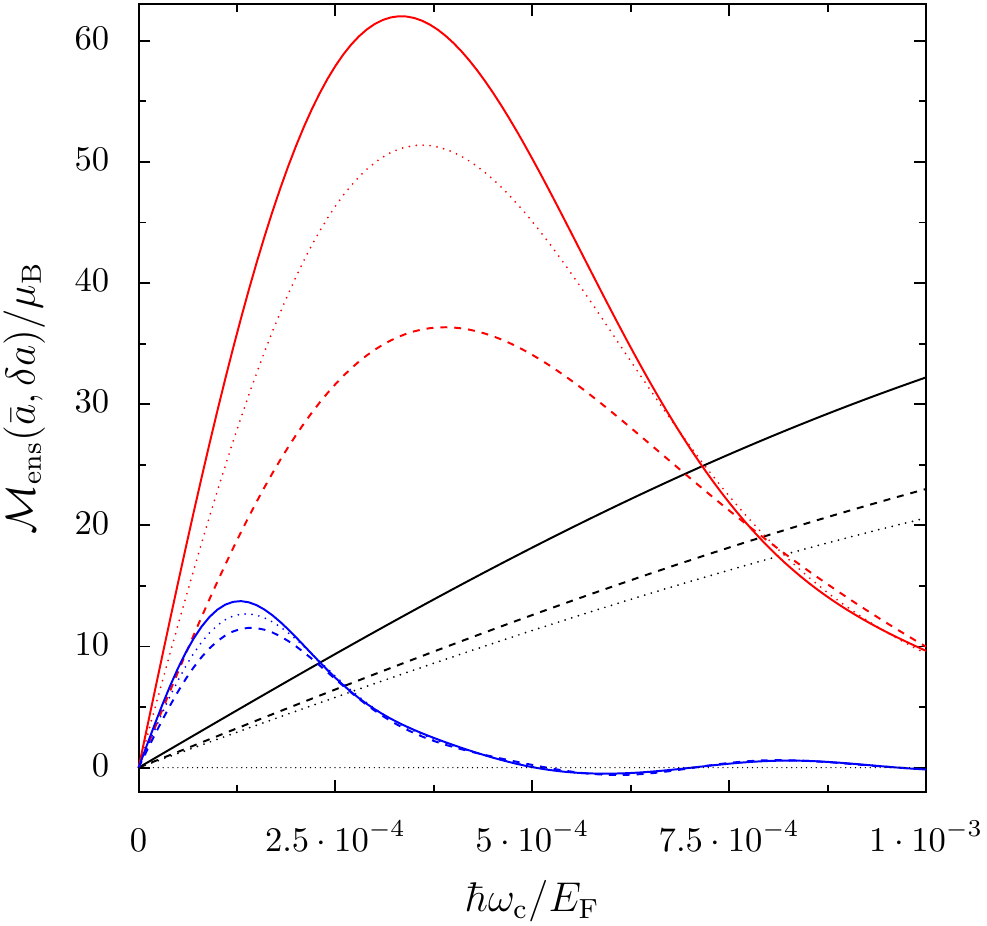}
\caption{\label{fig:M_ens}%
Ensemble-averaged magnetic moment per particle from Eq.~\eqref{eq:M_ens} as a 
function of magnetic field with mean radius 
$k_\mathrm{F}\bar a=20$ (black lines), 
$60$ (red lines), $100$ (blue lines),
and for size dispersions $k_\mathrm{F}\delta a=5$ (dashed lines) and $10$ 
(dotted lines). 
The diagonal contribution $\mathcal{M}_\mathrm{ens}^\mathrm{d}(\bar a)$ 
[cf.\ Eq.~\eqref{eq:M_scaling}] is shown as solid lines.
In the figure, $T/T_\mathrm{F}=5\times 10^{-3}$.}
\end{figure}

The dependence on magnetic field of the ensemble-averaged magnetic moment 
per particle according to Eq.\ \eqref{eq:M_ens} is presented for various 
average radii and size dispersions in Fig.\ \ref{fig:M_ens}. 
The ensemble-averaged magnetic moment per nanoparticle can reach 
several tens of $\mu_\mathrm{B}$ for room temperature 
($T/T_\mathrm{F}=5\times10^{-3}$). 
Moreover, the behavior of $\mathcal{M}_\mathrm{ens}$ as a function of the 
applied magnetic field in a given interval depends significantly on the
average size of the ensemble. For the smallest size considered in 
Fig.~\ref{fig:M_ens} ($k_\mathrm{F}\bar{a}=20$, black lines), the magnetic 
moment increases monotonically with the magnetic field for the whole range 
of the parameter $\hbar\omega_\mathrm{c}/E_\mathrm{F}\propto H$ considered. 
For $k_\mathrm{F}\bar a=60$ (red lines), $\mathcal{M}_\mathrm{ens}$ becomes 
a decreasing function of the magnetic field after a critical value that 
depends on the size dispersion $\delta a$. 
For larger sizes ($k_\mathrm{F}\bar a=100$, blue lines), the previous 
nonmonotonic behavior appears at a smaller critical field, and eventually there 
occurs a sign inversion of $\mathcal{M}_\mathrm{ens}$ for even larger fields. 
In Sec.\ \ref{sec:discussion} we link these findings with the existing 
experimental results found in the literature. 

In the case $k_\mathrm{F}\delta a\gg1$, the exponential factor in 
Eqs.\ \eqref{eq:M_ens} and \eqref{eq:chi_ens} selects only the ``diagonal" 
subensemble of topological indexes for which $\nu=\nu'$ and $\eta=\eta'$. 
When applicable, such an approximation considerably simplifies the evaluation 
of the semiclassical expressions and allows for simple estimations of the 
zero-field susceptibility and the magnetic moment. 
The diagonal part of the magnetic susceptibility \eqref{eq:chi_ens} can be 
written as 
\begin{equation}
\label{eq:chi_scaling}
\frac{\chi_\mathrm{ens}^\mathrm{d}(\bar a)}{|\chi_\mathrm{L}|}=
18\pi k_\mathrm{F}\bar{a} 
\sum_{\substack{\nu>0\\\eta>2\nu}}
\mathcal{F}_{\nu\eta}^{\nu\eta}R^2(L_{\nu\eta}/L_T),
\end{equation}
which is  positive since 
$\mathcal{F}_{\nu\eta}^{\nu\eta}>0$ [cf.\ Eq.\ \eqref{eq:f}].
As can be seen in Fig.\ \ref{fig:chi_ens}, this diagonal contribution 
(black solid line) provides a good account of the behavior of 
$\chi_\mathrm{ens}$ for large $k_\mathrm{F}\delta a$.

\begin{figure}[tb]
\includegraphics[width=\linewidth]{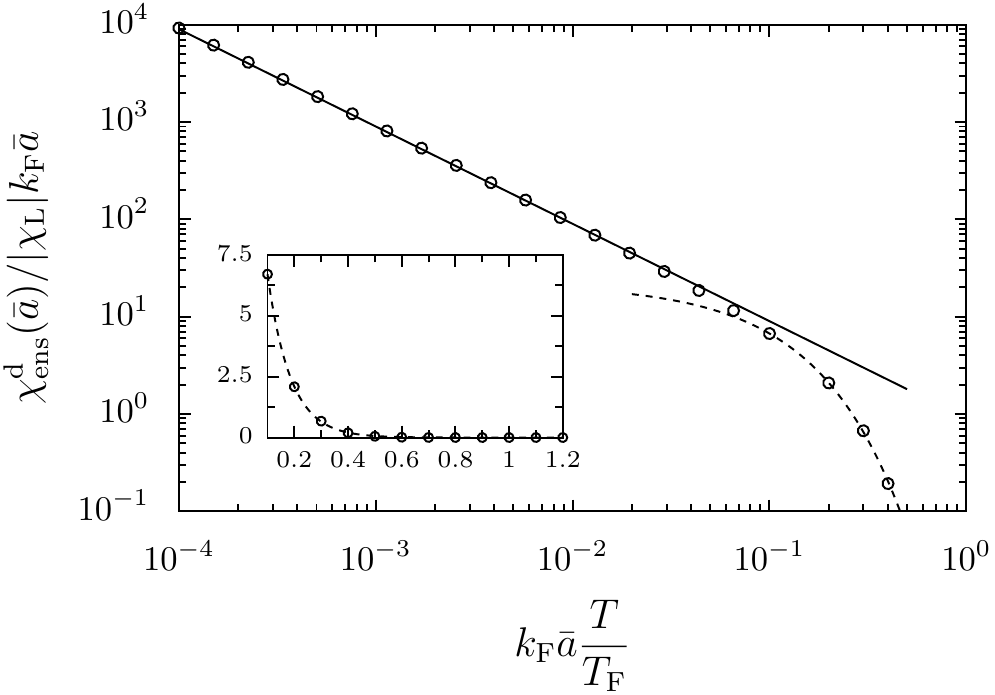}
\caption{\label{fig:chi_ens_T}%
Log-log plot of the diagonal contribution $\chi_\mathrm{ens}^\mathrm{d}(\bar{a})$ 
to the ensemble-averaged zero-field susceptibility (in units of 
$|\chi_\mathrm{L}|$) scaled with $k_\mathrm{F}\bar{a}$ as a function of 
$k_\mathrm{F}\bar{a}\frac{T}{T_\mathrm{F}}$.
Circles: numerical evaluation of Eq.\ \eqref{eq:chi_scaling}. 
Solid line: Curie-type law \eqref{eq:Curie}, with $C$ as given in 
Eq.~\eqref{eq:C}.
Dashed line: exponential fit \eqref{eq:exp_fit}, with $c_1\simeq22$ and 
$c_2\simeq12$.
Inset: Linear-scale plot of the same data, but for a different range of 
$k_\mathrm{F}\bar{a}\frac{T}{T_\mathrm{F}}$.}
\end{figure}

Interestingly, Eq.\ \eqref{eq:chi_scaling} is a function of the single 
parameter $k_\mathrm{F}\bar{a} \frac{T}{T_\mathrm{F}}=2\bar a/\pi L_T$ 
when scaled with $k_\mathrm{F}\bar{a}$. 
This can be seen from the argument of the thermal function \eqref{eq:R_T},
$L_{\nu\eta}/L_T=\pi\eta\sin{\varphi_{\nu\eta}}k_\mathrm{F}\bar{a}\frac{T}{T_F}$, 
and is exemplified in Fig.\ \ref{fig:chi_ens_T}, where the circles correspond 
to a numerical evaluation of the sum over the topological indexes in 
Eq.\ \eqref{eq:chi_scaling}.
Remarkably, for $k_\mathrm{F}\bar a\frac{T}{T_\mathrm{F}}\ll1$ 
(with $k_\mathrm{F}\bar a\gg1$), Eq.\ \eqref{eq:chi_scaling} follows the 
Curie-type law 
\begin{equation}
\label{eq:Curie}
\frac{\chi_\mathrm{ens}^\mathrm{d}}{|\chi_\mathrm{L}|}=\frac{C}{T/T_\mathrm{F}}, 
\end{equation} 
\textit{independent} of the average size $\bar a$ of the
nanoparticles. 

The prefactor $C$ of the above Curie law can be analytically evaluated along 
the lines leading to the semiclassical result \eqref{eq:chi_1_SC} 
and presented in Appendix~\ref{app:Curie}. 
First, the thermal factor (squared) in Eq.\ \eqref{eq:chi_scaling} is replaced 
by a Heaviside step function which cuts trajectories longer than 
$L_\mathrm{max}=\alpha L_T$ ($\alpha\simeq1.6$, see Sec.\ \ref{sec:1NP}). 
Second, the sums over the topological indexes are approximately evaluated by 
replacing them by integrals. 
To leading order in $k_\mathrm{F}\bar a\frac{T}{T_\mathrm{F}}\ll1$, 
we then obtain
\begin{equation}
\label{eq:C}
C=\frac{9\alpha}{16}.
\end{equation}
The result \eqref{eq:Curie}, together with Eq.\ \eqref{eq:C}, is shown by the 
solid line in Fig.\ \ref{fig:chi_ens_T}. As can be seen from the main 
figure, there is excellent quantitative agreement between the numerical 
evaluation of Eq.\ \eqref{eq:chi_scaling} (circles) and the approximate result 
\eqref{eq:Curie} (solid line) for small nanoparticle sizes and/or low 
temperatures. 

For larger values of the parameter $k_\mathrm{F}\bar{a}\frac{T}{T_\mathrm{F}}$, 
the susceptibility resulting from Eq.\ \eqref{eq:chi_scaling} deviates from the 
Curie-type law and is exponentially suppressed with temperature. 
It can be fitted by 
\begin{equation}
\label{eq:exp_fit}
\frac{\chi_\mathrm{ens}^\mathrm{d}(\bar a)}{|\chi_\mathrm{L}|}
=c_1 k_\mathrm{F}\bar{a}
\exp{\left(-c_2 k_\mathrm{F}\bar{a}\frac{T}{T_\mathrm{F}}\right)}, 
\end{equation} 
with $c_1\simeq22$ and $c_2\simeq 12$. 
Such a behavior can be traced back to the exponential suppression induced by 
the thermal factor \eqref{eq:R_T} even for the shortest trajectories, in line 
with our discussion of the high-temperature regime for $\chi^{(1)}$ in 
Sec.\ \ref{sec:1NP}.

\begin{figure}[tb]
\includegraphics[width=\linewidth]{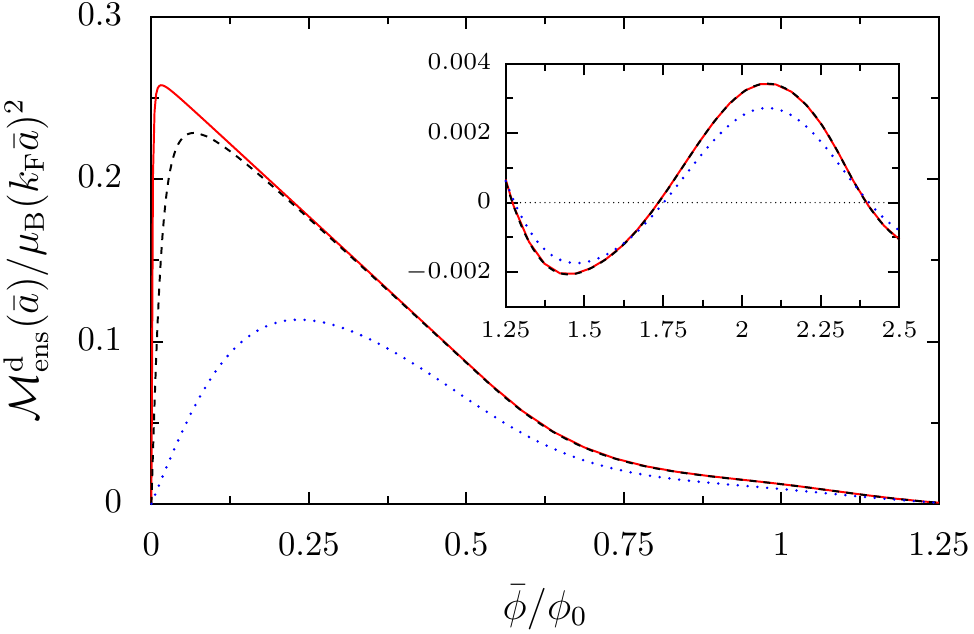}
\caption{\label{fig:M_ens_T}%
Diagonal contribution to the ensemble-averaged magnetic moment from 
Eq.\ \eqref{eq:M_scaling} scaled with $(k_\mathrm{F}\bar{a})^2$ as a function 
of the reduced magnetic flux $\bar{\phi}/\phi_0$ for 
$k_\mathrm{F}\bar{a}\frac{T}{T_\mathrm{F}}=10^{-3}$ (red solid line), 
$10^{-2}$ (black dashed line), and $10^{-1}$ (blue dotted line). 
Inset: Same as the main figure for a different range of $\bar{\phi}/\phi_0$.}
\end{figure}
Similarly to the case of the zero-field susceptibility, we consider the 
diagonal contribution [terms with $\nu=\nu'$ and $\eta=\eta'$ in 
Eq.\ \eqref{eq:M_ens}]
\begin{align}
\label{eq:M_scaling}
\frac{\mathcal{M}_\mathrm{ens}^\mathrm{d}(\bar a)}{\mu_\mathrm{B}}=&\,
-12(k_\mathrm{F}\bar a)^2
\sum_{\substack{\nu>0\\\eta>2\nu}}
\frac{\mathcal{F}_{\nu\eta}^{\nu\eta}}
{\eta\cos{\varphi_{\nu\eta}}\sin{\varphi_{\nu\eta}}}
\nonumber\\
&\times
R^2(L_{\nu\eta}/L_T) j'_0\left(2\pi\phi_{\nu\eta}/\phi_0\right)
j_0\left(2\pi\phi_{\nu\eta}/\phi_0\right)
\end{align}
to the magnetic moment per nanoparticle, which becomes dominant in the case 
$k_\mathrm{F}\delta a\gg1$ (solid lines in Fig.\ \ref{fig:M_ens}). 
Once scaled with $(k_\mathrm{F}\bar{a})^2$, Eq.\ \eqref{eq:M_scaling} only 
depends on the two following parameters: (i) the normalized flux
$\bar{\phi}/\phi_0$ appearing in the argument of the spherical Bessel function 
and its derivative in Eq.\ \eqref{eq:M_scaling} ($\bar{\phi}=\pi\bar{a}^2H$ is 
the average magnetic flux through a nanoparticle); 
(ii) the ratio between average radius and thermal length 
$2\bar{a}/\pi L_T=k_\mathrm{F}\bar{a}\frac{T}{T_\mathrm{F}}$ through the 
argument of the thermal reduction factor \eqref{eq:R_T}. 
Figure \ref{fig:M_ens_T} presents the flux dependence of the diagonal 
contribution \eqref{eq:M_scaling} scaled with $(k_\mathrm{F}\bar{a})^2$.
For weak flux, $\bar{\phi}\ll\phi_0$, the magnetic moment increases linearly 
with magnetic field and its temperature dependence follows a Curie-type law as 
shown for the susceptibility [see Eq.~\eqref{eq:Curie}].
For larger flux, a maximal value is attained and 
$\mathcal{M}_\mathrm{ens}^\mathrm{d}$ decreases until it reaches negative values 
and oscillates as a function of flux, resembling the de Haas-van Alphen
oscillations~\cite{landau_statphys} that would occur for much larger magnetic 
flux.
As the temperature decreases, the magnetic moment increases significantly
at weak magnetic field, reaching very high values.

\section{Magnetic response of individual nanoparticles}
\label{sec:ind}
In the previous section we discussed the situation of a nanoparticle ensemble, 
which is the case were the magnetic response has been abundantly measured. 
The magnetic response of an individual nanoparticle, given by
\begin{equation}
\mathcal{M}=\mathcal{M}^{(1)}+\mathcal{M}^{(2)}
\end{equation}
and 
\begin{equation}
\label{eq:chi_tot}
\chi=\chi^{(1)}+\chi^{(2)},
\end{equation}
has considerable interest for two reasons. 
Firstly, $\mathcal{M}$ and $\chi$ become relevant when analyzing the 
experimental conditions aiming at measurements on relatively small numbers of 
particles or in the case of single nanoparticles. 
These conditions could be achieved, e.g., using magnetic force 
microscopy~\cite{saenz87_JAP, donni2017} of nanoparticles deposited on a 
nonmagnetic substrate. 
Secondly, as we discuss in Appendix \ref{app:int}, if interactions among the 
nanoparticles of the ensemble are included in the description, the 
single-particle magnetic moment $\mathcal{M}$ becomes a crucial ingredient
of the model describing the magnetic response 
of coupled nanoparticles. 

\begin{figure}[tb]
\includegraphics[width=\linewidth]{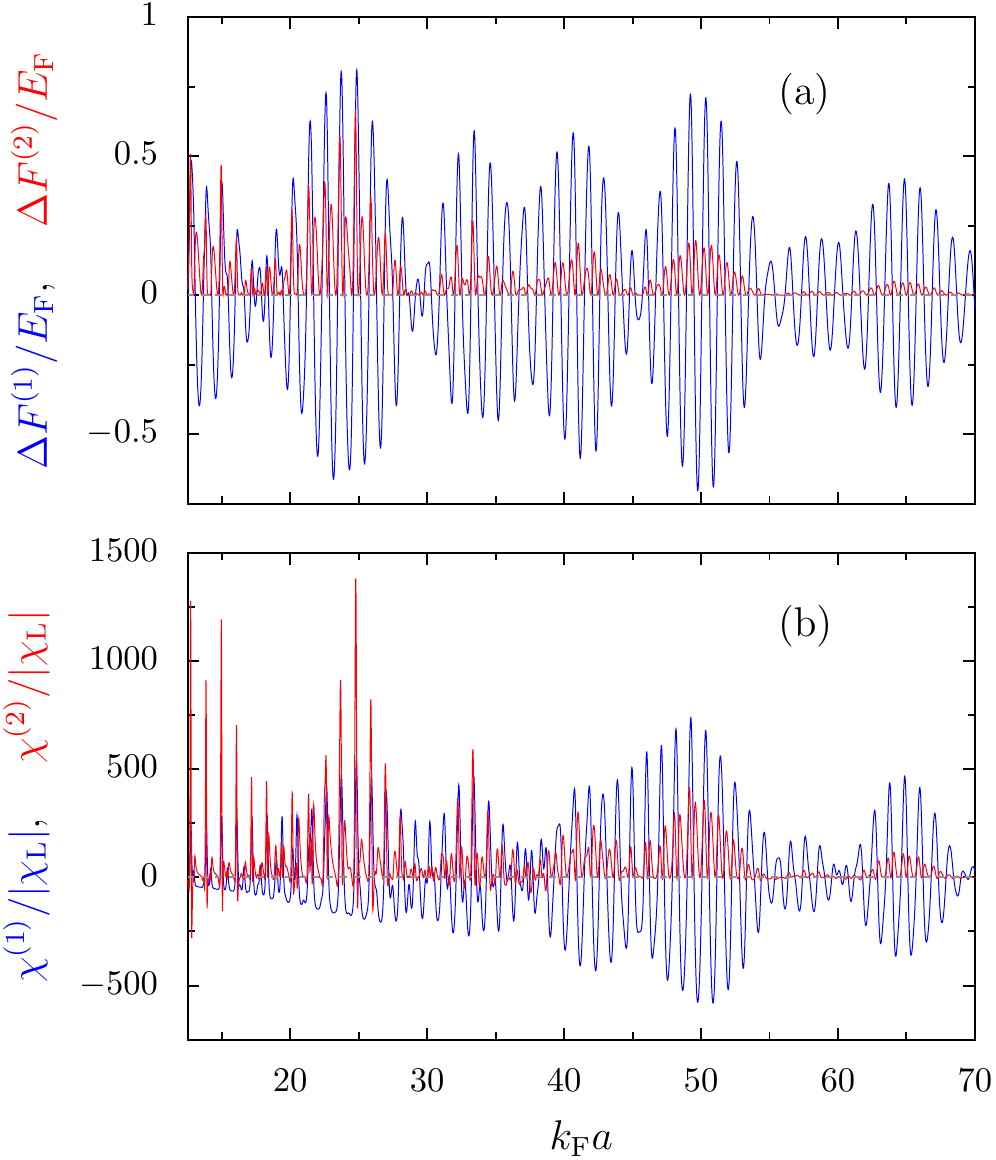}
\caption{\label{fig:chi_2}%
Blue lines: grand-canonical (a) free energy $\Delta F^{(1)}$ at $H=0$ 
(scaled with the Fermi energy $E_\mathrm{F}$) from Eq.\ \eqref{eq:F1_result} 
at a temperature $T/T_\mathrm{F}=5\times10^{-3}$ and (b) corresponding 
zero-field susceptibility $\chi^{(1)}$ from Eq.\ \eqref{eq:chi_1} 
(cf.\ blue line in Fig.\ \ref{fig:chi_1NP}) as a function of $k_\mathrm{F}a$. 
Red lines: canonical contribution $\Delta F^{(2)}$ from 
Eq.\ \eqref{eq:F2_result} [panel (a)] and zero-field susceptibility 
$\chi^{(2)}$ from Eq.\ \eqref{eq:chi_2} [panel (b)].}
\end{figure}

As discussed in Sec.\ \ref{sec:formalism}, the fulfillment of the condition 
$\Delta F^{(2)}\ll |\Delta F^{(1)}|$, at the basis of our semiclassical
thermodynamic formalism, depends on the values of $k_\mathrm{F}a$ and 
$T/T_\mathrm{F}$. 
In order to quantify these constraints, we present in Fig.\ \ref{fig:chi_2}(a) 
[Fig.\ \ref{fig:chi_2}(b)] the values of $\Delta F^{(1)}$ [$\chi^{(1)}$] in blue, 
and $\Delta F^{(2)}$ [$\chi^{(2)}$] in red, for room temperature 
($T/T_\mathrm{F}=5\times10^{-3}$) and a reduced $k_\mathrm{F}a$ span 
as compared to the one shown in Fig.\ \ref{fig:chi_1NP}. 
At the lowest values considered for $k_\mathrm{F}a$, $\Delta F^{(2)}$ is 
comparable to $|\Delta F^{(1)}|$, but it rapidly becomes comparatively smaller 
for $k_\mathrm{F}a\gtrsim 30$ and then completely negligible for 
$k_\mathrm{F}a\gtrsim 50$. 
The semiclassical thermodynamic formalism is then applicable at room 
temperature over almost all the $k_\mathrm{F}a$ range, even if $|\chi^{(2)}|$ 
typically dominates $|\chi^{(1)}|$ up to $k_\mathrm{F}a\simeq30$. 
Consistent with these results, the magnetic moment $\mathcal{M}$ of an 
individual nanoparticle at room temperature is essentially given by 
$\mathcal{M}^{(1)}$ for the sizes shown in Fig.\ \ref{fig:M_1NP}, 
where $\mathcal{M}$ as a function of $k_\mathrm{F}a$ is indistinguishable 
from $\mathcal{M}^{(1)}$ on the scale of the figure (data not shown).

\begin{figure}[tb]
\includegraphics[width=\linewidth]{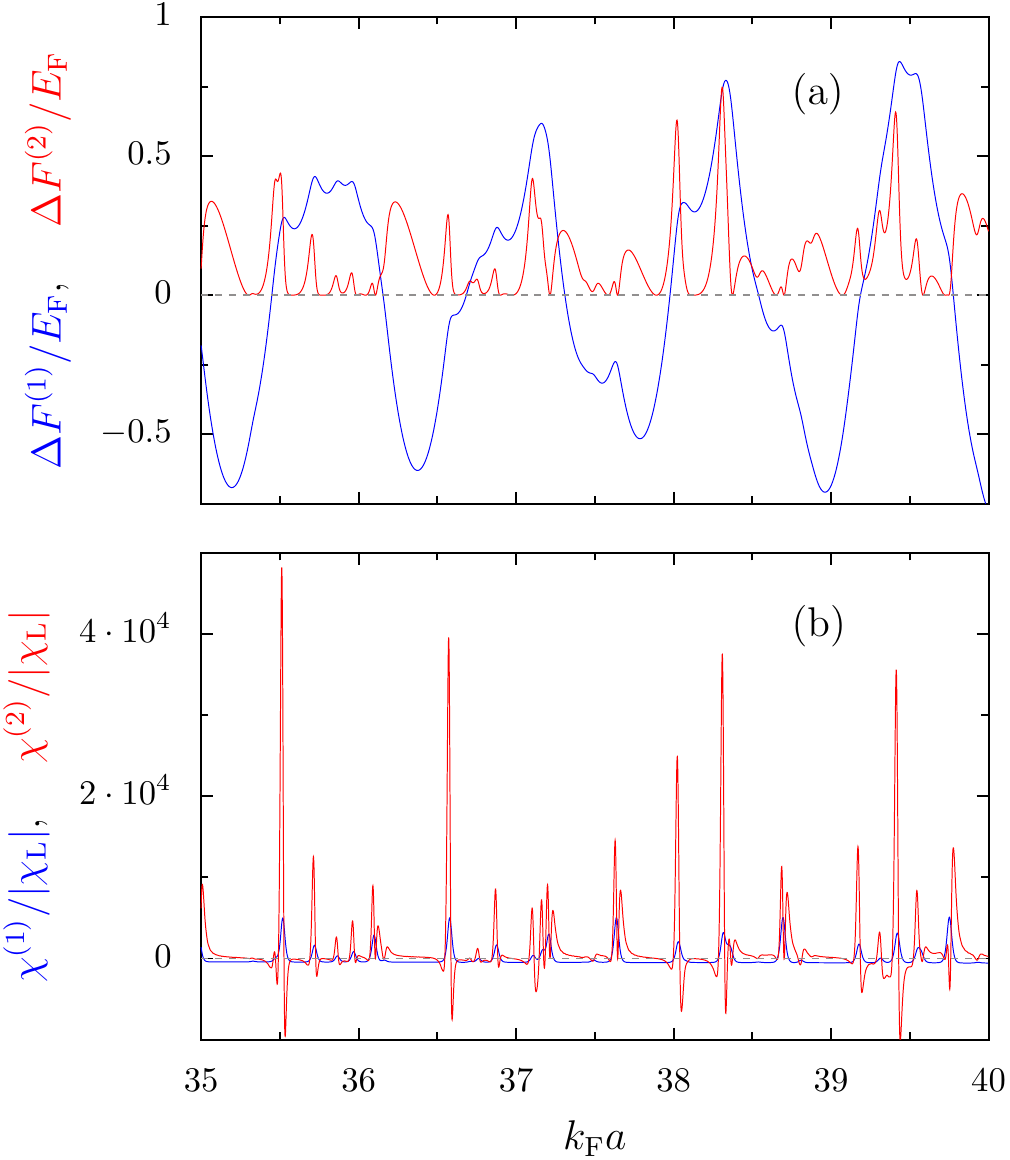}
\caption{\label{fig:chi_2_lowT}%
Low-temperature $T/T_\mathrm{F}=5\times10^{-4}$ results for the 
contributions to (a) the free energy and (b) the susceptibility. As in 
Fig.\ \ref{fig:chi_2}, blue lines represent the grand-canonical 
contributions from Eqs.\ \eqref{eq:F1_result} and \eqref{eq:chi_1}, 
and red lines depict the canonical contributions of 
Eqs.\ \eqref{eq:F2_result} and \eqref{eq:chi_2}.}
\end{figure}

The situation at low temperature $T/T_\mathrm{F}=5\times10^{-4}$ 
is presented in Fig.\ \ref{fig:chi_2_lowT} for a smaller range of 
$k_\mathrm{F}a$. 
Again, we can observe that the canonical contribution $\Delta F^{(1)}$ 
(blue line) is typically larger than the grand-canonical one 
$\Delta F^{(2)}$ (red line).
Even though the grand-canonical contribution to the susceptibility is 
larger that the canonical one, we expect the semiclassical formalism to 
yield at least qualitatively correct results for these and larger values 
of $k_\mathrm{F}a$.

\section{Discussion}
\label{sec:discussion}
The variety of possible magnetic responses (diamagnetic, paramagnetic, 
or ferromagnetic) experimentally observed calls for a systematic evaluation 
of the results yielded by different theoretical descriptions. 
Within our model presented in Sec.\ \ref{sec:model}, we obtained in 
Sec.\ \ref{sec:manyNPs} a paramagnetic response at weak fields for the case 
of an ensemble with a large number of noninteracting nanoparticles and a 
rather large size dispersion, as it is often the case in experiments. 
For increasing fields the magnetization of the nanoparticle ensemble could 
switch from its low-field paramagnetic behavior to a diamagnetic response 
(decreasing of the magnetization with the field and even a magnetization 
antialigned with the applied field, see Figs.\ \ref{fig:M_ens} and 
\ref{fig:M_ens_T}). 
While these changes are often observed in experiments \cite{nealo12_Nanoscale}, 
such behavior is usually interpreted as coming from spurious diamagnetic 
elements of the sample \cite{garci09_JAP}. 

In order to test the relevance of our approach, we will disregard the cases 
where the parameters of the sample are not completely known and exclude 
observations of ferromagnetism where, presumably, the interparticle 
interactions are important. 
We will thus concentrate on the experiments where the paramagnetic behavior 
has been clearly established. 

\begin{figure}[tb]
\includegraphics[width=\linewidth]{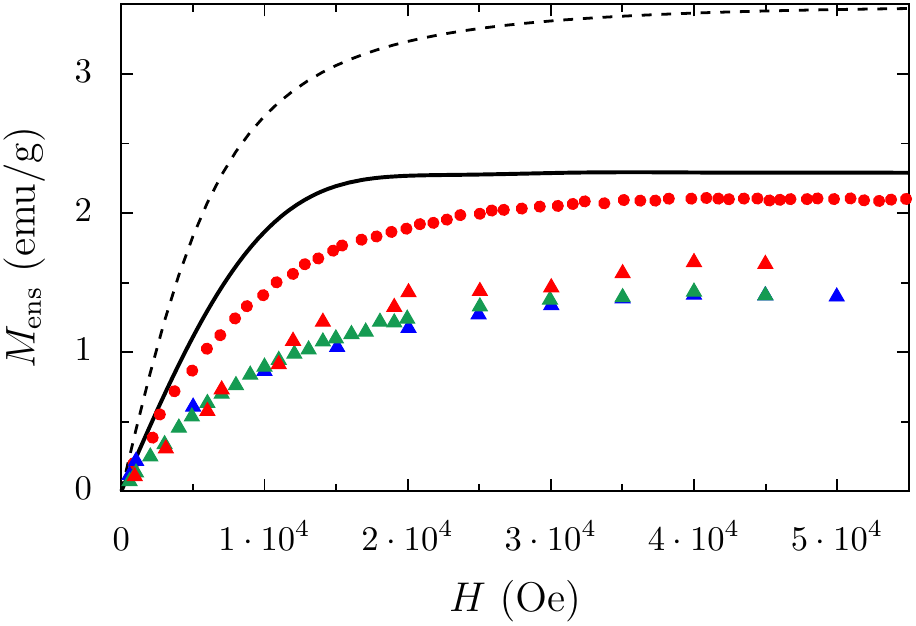}
\caption{\label{fig:experiments}%
Dots and triangles: measured magnetization $M_\mathrm{ens}$ of 
an ensemble of Au nanoparticles functionalized with various ligands (in 
electromagnetic units per gram of gold nanoparticles in the sample) as a 
function of applied field $H$ (in \oe{rsted}), with mean diameter 
$2\bar{a}=\unit[2.5]{nm}$, size dispersion $2\delta a=\unit[0.4]{nm}$, 
and at a temperature $T=\unit[1.8]{K}$. 
The data are taken from the experiments of 
Refs.\ \cite{hori99_JPA, nakae00_PhysicaB, hori04_PRB}. 
The corresponding ligands are: polyvinyl pyrolidone (PVP) [red dots
 (Refs.\ \cite{hori99_JPA, nakae00_PhysicaB}) and red triangles 
(Ref.~\cite{hori04_PRB})], polyacrylonitrile (PAN) [green triangles 
(Ref.~\cite{hori04_PRB})], and polyallyl amine hydrochloride (PAAHC)
[blue triangles (Ref.\ \cite{hori04_PRB})].
Solid line: nondiagonal magnetization from Eq.\ \eqref{eq:M_ens} with 
$2\delta a=\unit[0.4]{nm}$. 
Dashed line: diagonal approximation \eqref{eq:M_scaling}, corresponding to 
$2\delta a\rightarrow\infty$.}
\end{figure}

The pioneering experiments of Refs.\ \cite{hori99_JPA, nakae00_PhysicaB}, 
which also included palladium nanoparticles, have been extremely important 
in fostering the interest on the subject, by yielding large values of the 
saturation magnetic moment per nanoparticle (about $20\mu_\mathrm{B}$) in a 
regime where the magnetic interaction between the nanoparticles could be 
neglected. 
In Fig.~\ref{fig:experiments}, we reproduce the magnetization per gram of 
gold in the sample $M_\mathrm{ens}$ of Refs.~\cite{hori99_JPA, nakae00_PhysicaB} 
for gold nanoparticles surrounded by polyvinyl pyrolidone (PVP) ligands 
(red dots) having a mean diameter $2\bar a\simeq\unit[2.5]{nm}$ and a relatively 
narrow size dispersion ($2\delta a\simeq\unit[0.4]{nm}$) at $T=\unit[1.8]{K}$. 
These experimental data, well represented by the Langevin function and exhibiting 
quasiparamagnetic field and temperature dependences, have been reproduced in 
different samples with various ligands (see triangles in 
Fig.~\ref{fig:experiments}), except in the case where strong covalent bonds 
get established with the nanoparticles~\cite{hori04_PRB}. 

The solid line in Fig.~\ref{fig:experiments} represents 
$\mathcal{M}_\mathrm{ens}/\varrho\mathcal{V}$, where $\mathcal{M}_\mathrm{ens}$ 
is given in Eq.\ \eqref{eq:M_ens} and $\varrho=\unit[19.3]{g/cm^3}$ is the mass 
density of gold, for the temperature, mean diameter, and width of the size 
distribution of the experimental data \cite{footnote:nondiag}. 
As no fitting parameters are invoked, the qualitative agreement between our 
theory and these sets of data makes us conclude that the orbital response 
is indeed a crucial ingredient in the cases where the nanoparticle interaction 
is negligible. 

It should be remarked that for the small values of $a$ and $T$ used in 
Fig.\ \ref{fig:experiments}, the semiclassical thermodynamic formalism becomes 
questionable. 
Notwithstanding, while in the formalism of Sec.\ \ref{sec:formalism} the 
temperature is the only parameter to smooth out the oscillations of the density 
of states of an individual nanoparticle, in an ensemble of nanoparticles there 
are other additional factors that contribute to smooth the density of states and 
then reduce the values of $\overline{\Delta F^{(2)}}$. 
Among them, the size dispersion characterized by $\mathcal{P}(a)$, 
the possibility of having deviations with respect 
to the perfectly spherical shape, and effects of structural or impurity-induced 
disorder. 
It is based on the latter effect that the canonical correction has been obtained 
for the problem of persistent currents in metallic nanostructures 
\cite{imry91, schmi91_PRL}.

For comparison purposes, we also present in Fig.\ \ref{fig:experiments} 
(dashed line) $M_\mathrm{ens}$ according to the diagonal approximation~\eqref{eq:M_scaling}. On the one hand, we see that the simple diagonal 
approximation is enough to provide a qualitative agreement with respect to the 
experimental data. 
On the other hand, we verify that the effect of the size dispersion $\delta a$, 
which is responsible for the difference between the two expressions, appears as 
a key element in achieving a quantitative agreement. 

The existing data yielding a paramagnetic zero-field susceptibility are 
more difficult to relate with the theoretical prediction of 
Figs.\ \ref{fig:chi_ens} and \ref{fig:chi_ens_T} 
and Eqs.\ \eqref{eq:chi_ens} and \eqref{eq:Curie}. 
While the value of $\chi$ that can be extracted from the magnetization curve of 
Refs.~\cite{hori99_JPA, nakae00_PhysicaB} is in qualitative agreement with 
Eq.\ \eqref{eq:Curie} and the reported zero-field susceptibility follows a 
clear Curie law, the numerical values are two orders of magnitude larger than 
the theoretical prediction. 
The inconsistency between the magnetization and susceptibility results of 
Refs.\ \cite{hori99_JPA, nakae00_PhysicaB} might be due to an incorrect 
handling of the units \cite{yamam_privatecomm}. 

The magnetization measurements of Yamamoto \textit{et al.}\ \cite{yamam04_PRL} 
yielded a paramagnetic susceptibility for an ensemble of gold nanoparticles 
with $2\bar a=\unit[1.9]{nm}$ and a log-normal size distribution. 
The reported susceptibility follows a Curie-type law, but with values which 
are several orders of magnitude smaller than the previously-discussed data or 
the theoretical curve of Fig.\ \ref{fig:chi_ens_T}, and it has been explained 
from the orbital moment of the Au $5d$ electrons. 

The susceptibility results of Bartolom\'e \textit{et al.}\ \cite{barto12_PRL} 
on gold nanoparticles with naturally thiol-containing protective agents, 
between $T=\unit[2.7]{K}$ and $\unit[10]{K}$, exhibit a paramagnetic response 
with a clear Curie law, but an order of magnitude smaller than 
the data of Refs.\ \cite{hori99_JPA, nakae00_PhysicaB}. 
The findings of Ref.~\cite{barto12_PRL} have been interpreted by invoking the 
holes of the Au $5d$ band induced by the thiols, and thus the comparison with 
our ligand-independent theoretical approach is problematic. 

Some of the reported ferromagnetic samples present an extremely narrow 
hysteresis loop \cite{guerr08_Nanotechnology, guerr08}, such that 
a quasiparamagnetic zero-field susceptibility can be inferred. 
The values thus obtained from the low-temperature data of 
Refs.\ \cite{guerr08_Nanotechnology, guerr08} result in a paramagnetic 
susceptibility which is one to two orders of magnitude smaller than our 
theoretical prediction, depending on the nature of the protective ligands. 

We thus conclude that the orbital magnetism contribution is always important 
for analyzing the cases yielding a paramagnetic response of an ensemble of 
nanoparticles. 
In the cases where the ligands do not considerably alter the electronic states 
of the isolated nanoparticles, a qualitative agreement between theory and 
experiment is obtained for the magnetization curves and in the fulfillment of 
a Curie-type law of the zero-field susceptibility for a large range of 
temperatures (up to about room temperature, for sufficiently small 
nanoparticles). 

The diamagnetic response obtained in some 
experiments~\cite{cresp04_PRL, dutta07_APL, guerr08_Nanotechnology, rhee13_PRL} 
can also be accounted for from the orbital magnetism, provided a narrow size 
dispersion or a peaked size distribution of the nanoparticles in the ensemble 
allow for the fluctuations of $\chi^{(1)}$ (see Figs.\ \ref{fig:chi_1NP} and 
\ref{fig:chi_1NP_lowT}) to dominate over the paramagnetic contribution of 
$\chi^{(2)}$. 
The speculation that a paramagnetic response of the ensemble could turn into a 
diamagnetic one under the influence of spin-orbit coupling \cite{rhee13_PRL}, 
in analogy with the sign inversion of the magnetoconductance~\cite{hikam80}, 
is invalidated by the theoretical result of Ref.~\cite{mathu91_PRB}.

The ferromagnetic results are not accountable from our model of noninteracting 
nanoparticles. 
However, as we show in Appendix \ref{app:int}, the semiclassical approach to 
orbital magnetism settles the basis of a rich interacting model that can be 
tackled by numerical calculations.

\section{Conclusion}
\label{sec:ccl}
We have investigated orbital magnetism in gold nanoparticles.
Specifically, we have considered spherical metallic particles in the 
jellium approximation and treated the electron-electron interactions 
within a mean-field approach. 
The orbital response of individual as well as ensembles of nanoparticles 
with a smooth size distribution have been calculated within a semiclassical 
formalism. 
While the magnetic response at weak fields of an individual nanoparticle can 
be anything from strongly diamagnetic to strongly paramagnetic depending
on its size, the ensemble-averaged response is always paramagnetic when 
neglecting the interparticle interactions. 
In particular, we have predicted that the ensemble-averaged zero-field 
susceptibility should present a Curie-type law at low temperature, 
independent of the average size of the nanoparticles. 
We have obtained a qualitative agreement with the existing experimental data 
on the magnetization of ensembles of diluted nanoparticles in the case where 
interparticle interactions are negligible and where the local modifications 
induced by the surrounding ligands are irrelevant. 
Our results do not depend on details of the electronic structure and are thus 
not limited to gold but can be applied to any spherically-symmetric metallic 
nanoparticles. 
Moreover, the proposed mechanism does not rely on organic ligands surrounding 
the particles. 

An important conclusion of our work is to counter the 
claim~\cite{yamam04_PRL, hori04_PRB} that the strong paramagnetic response of 
the nanoparticle ensemble constitutes a proof that the individual nanoparticles 
are ferromagnetic. 
Indeed, we have shown that the orbital response of a large nanoparticle 
ensemble with a relatively broad size distribution can attain a large 
paramagnetic value through the flux accumulation of the underlying classical 
trajectories. 

In order to obtain analytically-tractable results, we assumed that the 
nanoparticles are perfectly spherical. 
However, crystallographic faceting at the surface of the particles, as well as 
static impurities inside the clusters, would presumably render the underlying 
classical dynamics of the electrons chaotic. 
As is well
known~\cite{oppen94_PRB, ullmo95_PRL, richt96_PhysRep, richt96_PRB, richt96_JMP},
the orbital magnetism of classically-chaotic and/or disordered systems is less 
pronounced than that of purely integrable ones. 
The high values of magnetic susceptibilities we obtain should thus be taken 
with care when comparing our results with existing experiments using larger 
nanoparticles and/or when disorder becomes important. 
However, the qualitative trends we are predicting should not be affected by 
fine details of the electron dynamics. 

This work is an important step toward understanding the effect of orbital 
magnetism in assemblies of nanoparticles. While the results presented here may 
explain a tendency toward the low-field paramagnetic behavior of certain samples, 
two potentially important ingredients for fully understanding some experiments
reporting an anomalous magnetic behavior of gold nanoparticles 
have been put aside in this work, namely the interparticle magnetic dipolar 
interactions and a nonsmooth, peaked size distribution. 
The former may be necessary to obtain ferromagnetic behavior, as is observed 
in certain samples, and can in principle be addressed 
with the semiclassical tools developed in this paper within the model sketched 
in Appendix \ref{app:int}.
The latter might occur depending on the fabrication process due to shell 
effects~\cite{knigh84_PRL, heer93_RMP}. 
The size dispersion was shown to be a crucial factor in determining the magnetic 
response, and in the limit where it becomes so small as to represent a peaked 
size distribution, we no longer expect the vanishing of the contribution of 
$\chi^{(1)}$ upon the ensemble average. 
The resulting strong oscillation as a function of nanoparticle size could 
explain the variation in the observed behavior from strong paramagnetism to 
strong diamagnetism in macroscopically similar samples having very narrow 
size distributions. 
We hope that our work will motivate future experimental and theoretical 
work considering these challenging issues.

\begin{acknowledgments}
We thank Thierry Charitat, Bertrand Donnio, Jean-Louis Gallani, Cosimo Gorini, 
Jean-Paul Kappler, Christian M\'eny, Michel Orrit, Pierre Panissod, 
Mircea Rastei, and Fabrice Thalmann for stimulating discussions. 
We are indebted to Motohiro Suzuki, Toshiharu Teranishi, and Yoshiyuki Yamamoto 
for helpful correspondence. 
We acknowledge financial support from the 
French Agence Nationale de la Recherche (Project ANR-14-CE26-0005 Q-MetaMat).
\end{acknowledgments}


\appendix
\section{Exact and perturbative quantum calculations}
\label{app:quantum}
The semiclassical approach developed in this work is particularly useful in 
order to calculate the paramagnetic component of the magnetization and the 
zero-field susceptibility that determine the magnetic response of an 
ensemble of noninteracting nanoparticles. 
The contributions $\mathcal{M}^{(1)}$ and $\chi^{(1)}$ to the 
magnetic response of an individual nanoparticle can be accessed either through 
the semiclassical theory or through perturbative quantum calculations. 
It is therefore important to use the latter in order to establish a benchmark 
of the former and validate the use of semiclassics in the cases where the 
quantum calculations are too difficult to be implemented, like that of the 
average magnetization which necessitates to impose the condition of a constant 
number of electrons at finite temperature. 

To second order in the magnetic induction $B$, nondegenerate perturbation 
theory yields for the spectrum of the mean-field Hamiltonian
\eqref{eq:H_mf} the analytical result \cite{ruite91_PRL, leuwe93_PhD}
\begin{equation}
\label{eq:E_nlm}
E_{nlm_z}=E_{nl}^{(0)}+E_{nlm_z}^{(1)}+E_{nlm_z}^{(2)}.
\end{equation} 
Here, $E_{nl}^{(0)}=E_0 \zeta_{nl}^2$ are the eigenenergies of a zero-field 
sphere with infinite potential walls, where $\zeta_{nl}$ is the 
$n$\textsuperscript{th} zero of the spherical Bessel function $j_{l}(z)$, 
with $l$ the angular momentum quantum number. 
The first-order contribution corresponding to the paramagnetic term of the 
Hamiltonian \eqref{eq:H_mf} reads $E_{nlm_z}^{(1)}=\hbar\omega_\mathrm{c}m_z/2$ 
(in terms of the magnetic quantum number $m_z$), while the second-order 
correction (diamagnetic term) is 
$E_{nlm_z}^{(2)}=m\omega_\mathrm{c}^2a^2\mathcal{R}_{nl}\mathcal{Y}_l^{m_z}/8$, 
with 
\begin{equation}
\label{eq:R}
  \mathcal{R}_{nl}=\frac{1}{3}\left[1+\frac{(2l+3)(2l-1)}{2\zeta_{nl}^2}\right]
\end{equation}
and 
\begin{equation}
\label{eq:Y}
  \mathcal{Y}_l^{m_z}=1-\frac{1}{2l+1}
  \left[\frac{l^2-m_z^2}{2l-1}+\frac{(l+1)^2-m_z^2}{2l+3}\right].
\end{equation}

\begin{figure*}[tb]
\includegraphics[width=.75\linewidth]{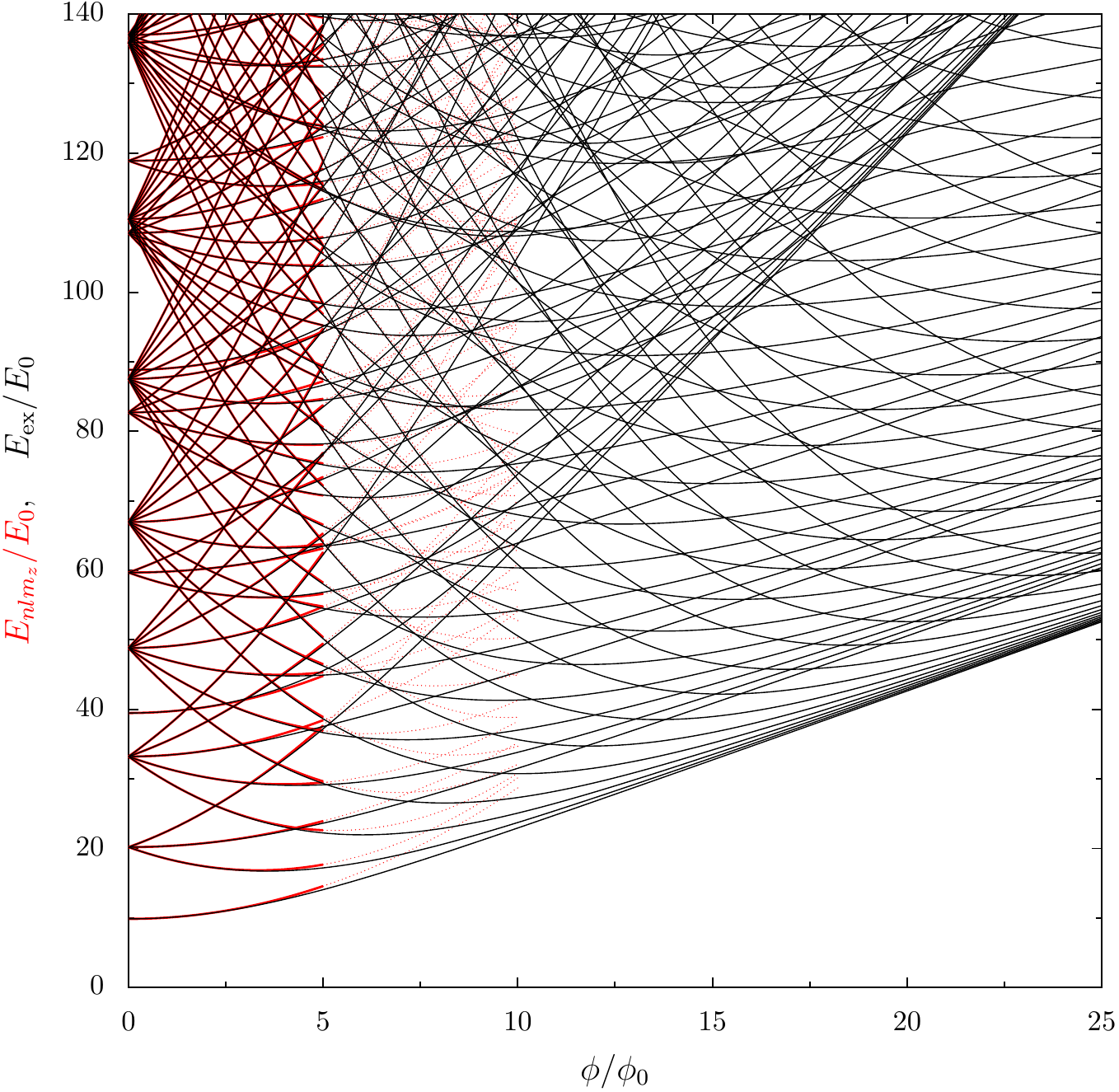}
\caption{\label{fig:exact_diag}%
Black lines: exact spectrum $E_\mathrm{ex}$ of the mean-field Hamiltonian 
\eqref{eq:H_mf} (scaled by $E_0=\hbar^2/2ma^2$) of a sphere as a function of 
the magnetic flux $\phi=\pi a^2 B$ in units of the flux quantum.
Red lines: perturbative spectrum $E_{nlm_z}$ from Eq.\ \eqref{eq:E_nlm}, 
showing the perturbative regime ($0<\phi/\phi_0\lesssim5$, solid red lines) 
and the region where perturbation theory starts to depart from the exact 
result ($5\lesssim\phi/\phi_0\lesssim10$, dotted red lines).}
\end{figure*}
In Fig.\ \ref{fig:exact_diag}, we compare the perturbative spectrum 
\eqref{eq:E_nlm} (red lines) for a given span of magnetic fields with
the exact spectrum $E_\mathrm{ex}$ resulting from a numerical diagonalization 
of the Hamiltonian~\eqref{eq:H_mf} (black lines). The magnetic fields needed to 
reach the regime of quantum Hall effect emerging at the right part of the plot 
are extremely high for the nanoparticles under consideration, but might be 
attainable for larger metallic nanoparticles or for semiconducting structures. 
The agreement between the perturbative and exact spectrum is very good up to 
magnetic fields corresponding to the (reduced) flux $\phi/\phi_0\simeq 5$, 
with $\phi=\pi a^2B$ (compare the solid red and black lines), while for 
larger fields, the perturbative energy levels (shown as dotted red lines) 
depart from the exact result. 
For the magnetic fields which we consider in the main text, the quantitative 
agreement is excellent, and the use of nondegenerate perturbation theory is 
appropriate since the perturbation does not break the cylindrical symmetry 
of the system. 

The quantum-mechanical magnetic moment $\mathcal{M}^{(1)}_\mathrm{q}$ and 
the corresponding zero-field susceptibility $\chi^{(1)}_\mathrm{q}$ can be 
readily obtained from the perturbative spectrum \eqref{eq:E_nlm} via the 
expressions 
$\mathcal{M}^{(1)}_\mathrm{q}=-2\sum_{nlm_z}(\partial_B E_{nlm_z}) f(E_{nlm_z})$ 
and 
$\chi^{(1)}_\mathrm{q}=(\partial_B \mathcal{M}^{(1)}_\mathrm{q})_{(B=0)}/\mathcal{V}$, 
where the factor of $2$ takes into account the spin degeneracy, yielding 
\cite{ruite91_PRL, leuwe93_PhD}
\begin{align}
\label{eq:M_1_QM}
    \frac{\mathcal{M}^{(1)}_\mathrm{q}}
    {\mu_\mathrm{B}}=&\;-2\sum_{n=1}^\infty\sum_{l=0}^\infty\sum_{m_z=-l}^{+l}
    f(E_{nlm_z})
    \nonumber\\
    &\times
    \left[m_z+\frac{(k_\mathrm{F} a)^2\hbar \omega_\mathrm{c}}{4E_\mathrm{F}}
    \mathcal{R}_{nl}\mathcal{Y}_{l}^{m_z}\right]
\end{align}
and
\begin{align}
\label{eq:chi_1_QM}
 \frac{\chi^{(1)}_\mathrm{q}}{|\chi_\mathrm{L}|}=&\;\frac{3\pi}{k_\mathrm{F}a}\sum_{n=1}^\infty\sum_{l=0}^\infty(2l+1)f(E_{nl}^{(0)})
 \nonumber\\
 &\times \left\{
 \left[1-f(E_{nl}^{(0)})\right]\frac{l(l+1)}{(k_\mathrm{F}a)^2T/T_\mathrm{F}}
- \mathcal{R}_{nl}
 \right\}.
\end{align}

The component $\mathcal{M}_\mathrm{q}^{(1)}$ arising from the perturbative 
spectrum \eqref{eq:E_nlm}, and shown in Fig.\ \ref{fig:exact_diag}, 
is presented in Fig.\ \ref{fig:M_1NP} (black lines), thus validating the 
use of semiclassical expansions at finite magnetic fields. 
We have checked that both the approximate quantum result $E_{nlm_z}$ and 
the exact diagonalization procedure yielding $E_\mathrm{ex}$ result in the 
same values for $\mathcal{M}^{(1)}_\mathrm{q}$.

\section{Semiclassical evaluation of the grand-canonical orbital magnetic susceptibility}
\label{app:sc}
In this appendix, we provide details of the semiclassical calculation of the 
grand-canonical orbital magnetic susceptibility leading 
to Eq.\ \eqref{eq:chi_1_SC} in the main text. 

\begin{figure}[tb]
\includegraphics[width=.7\linewidth]{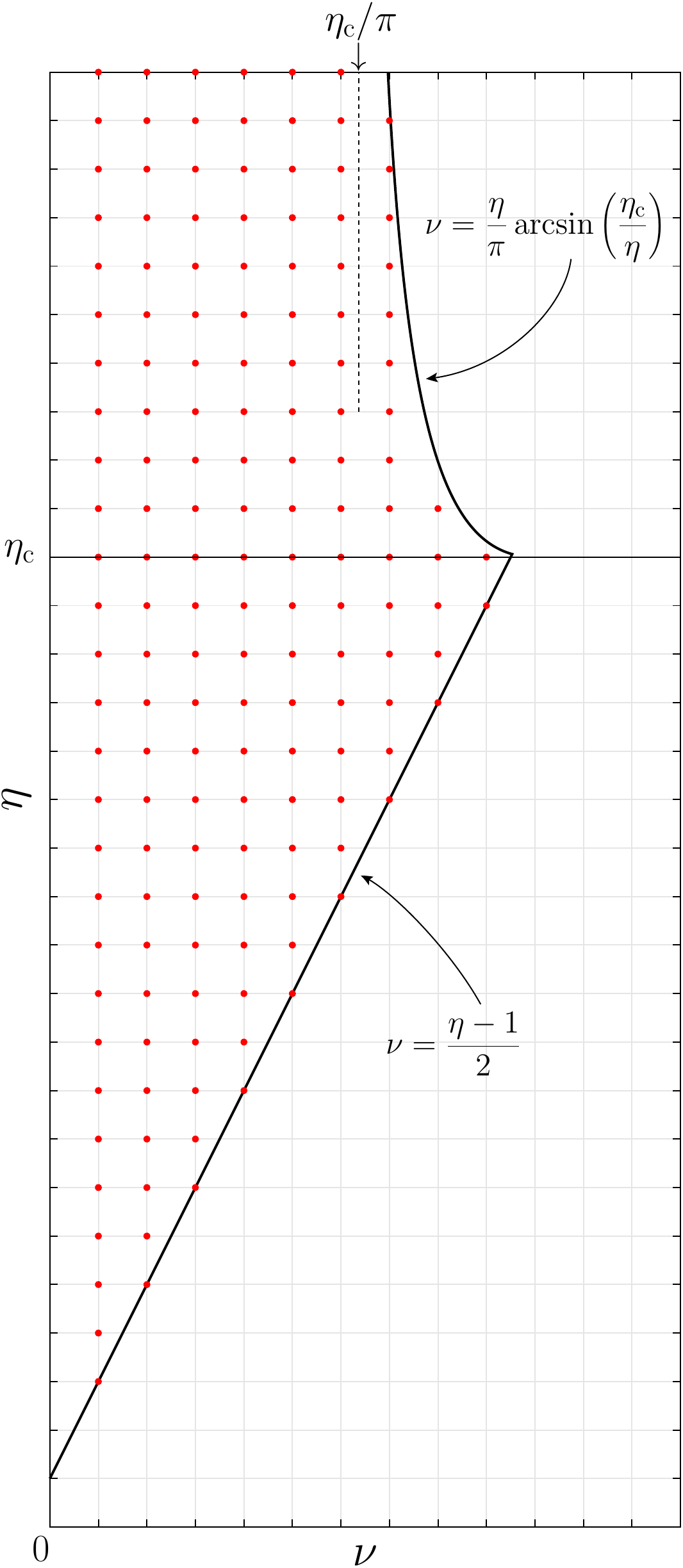}
\caption{\label{fig:shadow}%
Topological indexes $(\nu, \eta)$ contributing to the double sums in 
Eqs.\ \eqref{eq:chi_1_reordered} and \eqref{eq:chi_nu} (red dots).
The critical value $\eta_\mathrm{c}=\alpha L_T/2a$ separates the two 
summation regions with different values of $\nu_\mathrm{max}(\eta)$ given 
by the explicit formulas (solid lines). 
The dotted line depicts the limiting value of 
$\nu_\mathrm{max}$ when $\eta\rightarrow\infty$.}
\end{figure}

In the limit of low temperatures and/or small sizes 
($k_\mathrm{F}a\frac{T}{T_\mathrm{F}}\ll1$), we replace the thermal factor 
appearing in the semiclassical expansion \eqref{eq:chi_1} by a Heaviside 
step function that limits the contributing trajectories to the maximal 
length $L_\mathrm{max}=\alpha L_T$, yielding the condition on 
the topological index $\eta\leqslant\eta_\mathrm{c}$, with 
$\eta_\mathrm{c}=\alpha L_T/2a=(\alpha/\pi)(k_\mathrm{F}a
\frac{T}{T_\mathrm{F}})^{-1}\gg1$. 
Here, the parameter $\alpha\simeq1.6$ is chosen in such a way that the thermal 
factor \eqref{eq:R_T} presents the maximum derivative. 
Taking into account the above restriction and reordering the summations over 
$\nu$ and $\eta$ in Eq.\ \eqref{eq:chi_1} then lead to
\begin{align}
\label{eq:chi_1_reordered}
\frac{\chi^{(1)}}{|\chi_\mathrm{L}|}\simeq&\;6\sqrt{\pi}(k_\mathrm{F}a)^{3/2}
\sum_{\eta=3}^{\infty}
\sum_{\nu=1}^{\nu_\mathrm{max}
(\eta)}\mathrm{Re}\left\{\mathrm{e}^{-\mathrm{i}\theta_{\nu\eta}(k_\mathrm{F})}\right\}
\nonumber\\
&\times\frac{(-1)^\nu
\cos^3{\varphi_{\nu\eta}}\sin^{3/2}{\varphi_{\nu\eta}}}{\sqrt{\eta}}.
\end{align}
We have defined $\nu_\mathrm{max}(\eta)=\lfloor\frac{\eta-1}{2}\rfloor$ for $3\leqslant\eta\leqslant\lfloor\eta_\mathrm{c}\rfloor$ and 
$\nu_\mathrm{max}(\eta)=\lfloor\frac{\eta}{\pi}\arcsin{\left(\frac{\eta_\mathrm{c}}{\eta}\right)}\rfloor$ 
for $\eta\geqslant\lceil\eta_\mathrm{c}\rceil$.
The grid of points that represent the topological indexes $(\nu, \eta)$ 
contributing to the double sums of Eq.\ \eqref{eq:chi_1_reordered} are 
represented by red dots in Fig.~\ref{fig:shadow}.

The summation over the winding number $\nu$ in Eq.~\eqref{eq:chi_1_reordered} 
is then expressed using the Poisson summation formula, yielding
\begin{align}
\label{eq:chi_1_SP}
&\sum_{\nu=1}^{\nu_\mathrm{max}(\eta)}(-1)^\nu
\cos^3{\varphi_{\nu\eta}}
\sin^{3/2}{\varphi_{\nu\eta}}\
\mathrm{e}^{-\mathrm{i}\theta_{\nu\eta}(k_\mathrm{F})}
=\nonumber\\
&\sum_{l=-\infty}^{+\infty}\int_{1/2}^{\nu_\mathrm{max}(\eta)+1/2}\mathrm{d}\nu\
\cos^3{\varphi_{\nu\eta}}
\sin^{3/2}{\varphi_{\nu\eta}}\
\mathrm{e}^{\mathrm{i}\Psi_{\nu\eta,l}}, 
\end{align}
with the phase $\Psi_{\nu\eta,l}=(2l+1)\pi\nu-\theta_{\nu\eta}$. 
The above integral over $\nu$ is then performed using a stationary phase 
approximation that results in the stationary points 
$\bar{\nu}=(\eta/\pi)\arccos{([l+1/2]/k_\mathrm{F}a)}$.
Imposing that the latter belong to the $\nu$ integration interval in 
Eq.\ \eqref{eq:chi_1_SP} gives the following restriction over the index $l$: 
\begin{equation}
\label{eq:condition}
2k_\mathrm{F}a\cos{\left(\frac{\pi}{\eta}\left[\nu_\mathrm{max}(\eta)+
\frac 12\right]\right)}\leqslant 2l
+1\leqslant2k_\mathrm{F}a\cos{\left(\frac{\pi}{2\eta}\right)}.
\end{equation}
Substituting $j=2l+1$ in Eq.\ \eqref{eq:chi_1_SP} thus leads to
\begin{align}
\label{eq:chi_1_inter}
&\sum_{\nu=1}^{\nu_\mathrm{max}(\eta)}(-1)^\nu
\cos^3{\varphi_{\nu\eta}}
\sin^{3/2}{\varphi_{\nu\eta}}\
\mathrm{e}^{-\mathrm{i}\theta_{\nu\eta}(k_\mathrm{F})}
=\nonumber\\
&\sum_{\mathrm{odd}\ j}\left(\frac{j}{2k_\mathrm{F}a}\right)^3 
\sqrt{1-\left(\frac{j}{2k_\mathrm{F}a}\right)^2}
\sqrt{\frac{\eta}{\pi k_\mathrm{F}a}}\
\mathrm{e}^{-\mathrm{i}\eta S_j}, 
\end{align}
where the reduced radial action $S_j$ is defined in Eq.\ \eqref{eq:S_j} 
and where the summation over the odd integer $j$ is restricted by the 
condition \eqref{eq:condition}.
Incorporating the result \eqref{eq:chi_1_inter} into 
Eq.\ \eqref{eq:chi_1_reordered} then yields Eq.\ \eqref{eq:chi_1_SC}.

\section{Derivation of the Curie-type law for ensembles of noninteracting nanoparticles}
\label{app:Curie}
In this appendix, we demonstrate the Curie-type law for the orbital magnetic 
susceptibility of noninteracting ensembles of metallic nanoparticles
that arises at low temperature and/or for small sizes 
[cf.\ Eq.\ \eqref{eq:Curie} in the main text]. 
Here and in what follows, we adopt the notation of Appendix \ref{app:sc}, 
with the modification of changing the individual nanoparticle radius $a$ by 
the average radius $\bar{a}$ of the ensemble. 

Starting from Eq.\ \eqref{eq:chi_scaling}, in the limit 
$k_\mathrm{F}\bar{a}\frac{T}{T_\mathrm{F}}\ll1$ we replace the thermal 
factor squared by a Heaviside step function which cuts trajectories 
longer than $L_\mathrm{max}=\alpha L_T$, leading to
\begin{equation}
\label{eq:chi_nu}
\frac{\chi_\mathrm{ens}^\mathrm{d}}{|\chi_\mathrm{L}|}\simeq 
18\pi k_\mathrm{F}\bar{a}
\sum_{\eta=3}^{\infty}
\sum_{\nu=1}^{\nu_\mathrm{max}(\eta)}
\mathcal{F}_{\nu\eta}^{\nu\eta}, 
\end{equation}
with $\mathcal{F}_{\nu\eta}^{\nu\eta}=\frac{1}{\eta}\cos^4{\varphi_{\mu\eta}}\sin^3{\varphi_{\mu\eta}}$ 
[cf.\ Eq.\ \eqref{eq:f}]. 
Like in the case of Appendix \ref{app:sc}, the grid $(\nu, \eta)$ of 
points contributing to the double sums of Eq.\ \eqref{eq:chi_nu}
are represented by red dots in Fig.~\ref{fig:shadow}.
Since $\mathcal{F}_{\nu\eta}^{\nu\eta}$ has a smooth dependence on $\nu$, 
we approximate the summation over $\nu$ in Eq.\ \eqref{eq:chi_nu} by an 
integral, leading to
\begin{align}
\sum_{\nu=1}^{\nu_\mathrm{max}(\eta)}\mathcal{F}_{\nu\eta}^{\nu\eta}\simeq&\;
\frac{1}{\pi}
\left\{
\frac{1}{5}\left[\cos^5{\left(
\frac{\pi}{\eta}\right)}-\cos^5{\left(\frac{\pi\nu_\mathrm{max}(\eta)}{\eta} \right)}\right]
\right.
\nonumber\\
&-\left.
\frac{1}{7}\left[\cos^7{\left(\frac{\pi}{\eta}\right)}-\cos^7{\left(\frac{\pi\nu_\mathrm{max}(\eta)}{\eta}\right)}\right]
\right\}.
\end{align}

In Eq.\ \eqref{eq:chi_nu} the summation over $\eta$ is dominated by relatively 
large values of $\eta$. 
Therefore we make the approximation 
$\cos^5{({\pi}/{\eta})}\approx\cos^7{({\pi}/{\eta})}\approx1$ in the 
expression above. 
Moreover, for $\eta\leqslant\lfloor\eta_\mathrm{c}\rfloor$, we have 
$\cos{\big(\pi\nu_\mathrm{max}(\eta)/{\eta}\big)}\simeq \sin{(\pi/2\eta)}$, 
so that $\cos^5{\big({\pi\nu_\mathrm{max}(\eta)}/{\eta}\big)}\approx\cos^7{\big({\pi\nu_\mathrm{max}(\eta)}/{\eta}\big)}\approx0$,  
while for $\eta\geqslant\lceil\eta_\mathrm{c}\rceil$, we have $\cos{\big({\pi\nu_\mathrm{max}(\eta)}/{\eta}\big)}=[1-(\eta_\mathrm{c}/\eta)^2]^{1/2}$. 
Thus, Eq.\ \eqref{eq:chi_nu} yields 
\begin{align}
\frac{\chi_\mathrm{ens}^\mathrm{d}}{|\chi_\mathrm{L}|}\simeq&\; 18\pi k_\mathrm{F}\bar{a} 
\Bigg\{
\frac{2\eta_\mathrm{c}}{35}
\nonumber\\
&+
\int_{\eta_\mathrm{c}}^\infty\mathrm{d}\eta
\left[
\frac 15 \left(1-\left[1-\left(\frac{\eta_\mathrm{c}}{\eta}\right)^2\right]^{5/2}\right)
\right.
\nonumber\\
&-\left.\frac 17 \left(1-\left[1-\left(\frac{\eta_\mathrm{c}}{\eta}\right)^2\right]^{7/2}\right)
\right]
\Bigg\}
\end{align}
in the limit $\eta_\mathrm{c}\gg1$. Performing the remaining $\eta$ integral, 
we find the Curie-type law \eqref{eq:Curie}, with the prefactor $C$ as given 
in Eq.\ \eqref{eq:C}.

\section{Interacting model}
\label{app:int}
Throughout this work we have neglected the magnetic dipolar interactions 
between the different nanoparticles composing the nanoparticle ensemble 
encountered in the existing experiments. 
While such a hypothesis seems reasonable for fairly diluted samples, 
the appearance of a macroscopic ferromagnetic response in certain samples 
indicates that interparticle interactions might be at play.

Disregarding the possibility that individual nanoparticles are themselves 
ferromagnetic, and then possess a permanent magnetic moment, the 
single-particle magnetic moment arising from the orbital motion only exists 
at nonvanishing magnetic fields. 
The formalism developed in this work (see Secs.~\ref{sec:formalism} and 
\ref{sec:1NP}) allows us to write the macroscopic magnetization of an ensemble of 
$\mathcal{N}$ nanoparticles as
\begin{equation}
\label{eq:Mtot}
\mathbf{M}=\frac{\sum_{i=1}^\mathcal{N}\boldsymbol{\mathcal{M}}_i}{\sum_{i=1}^\mathcal{N}\mathcal{V}_i}.
\end{equation}
We note $\boldsymbol{\mathcal{M}}_i$ and $\mathcal{V}_i$ the magnetic moment 
and the volume, respectively, of the individual nanoparticles. 
The orientation of the magnetizations of different nanoparticles may 
be different once interactions are present. We thus have to 
consider the magnetization vector $\boldsymbol{\mathcal{M}}_i$ here.
The effective field $\mathbf{H}_i$ acting on nanoparticle $i$ is given by 
the external field $\mathbf{H}$ and the contributions generated by the other 
nanoparticles, that is
\begin{equation}
\label{eq:B_i}
\mathbf{H}_i=\mathbf{H}+\sum_{\substack{j=1\\(j\neq i)}}^\mathcal{N}
\frac{3\hat{r}_{ij}\left(\hat{r}_{ij}\cdot\boldsymbol{\mathcal{M}}_j\right)-
\boldsymbol{\mathcal{M}}_j}{r_{ij}^3}, 
\end{equation}
where $\hat{r}_{ij}$ is the unit vector in the direction linking 
nanoparticles $i$ and $j$, separated by the distance $r_{ij}$.
Assuming sufficiently weak internal fields, such that 
$\boldsymbol{\mathcal{M}}_i=\mathcal{V}_i\chi_i\mathbf{H}_i$, 
with $\chi_i$ the size-dependent orbital susceptibility of nanoparticle $i$ 
given by Eq.\ \eqref{eq:chi_tot}, Eqs.\ \eqref{eq:Mtot} and \eqref{eq:B_i} 
give rise to an extremely involved selfconsistent problem.

The above model includes disorder, through the random positions of the 
nanoparticles, and frustration, through the highly oscillating function 
$\chi_i$ that depends on the nanoparticle size, which are the two ingredients 
characterizing the rich physics at play in spin glasses \cite{mezard, fischer}. 
However, the problem at hand has two features that make its treatment even more 
difficult: the genuinely long-ranged nature of the magnetic dipolar 
interparticle interactions and the absence of a permanent magnetic moment of 
the nanoparticles. 
Numerical simulations, beyond the scope of the present paper, need to be 
developed by taking special care to the finite-size effects and to the 
contribution $\chi^{(2)}$ to the highly oscillating zero-field 
susceptibility of the individual nanoparticles.



\begin{thebibliography}{}

\bibitem{kubo62_JPSJ}
R. Kubo, 
Electronic properties of metallic fine particles. I., 
\href{http://dx.doi.org/10.1143/JPSJ.17.975}
{J. Phys. Soc. Jpn. \textbf{17}, 975 (1962).}

\bibitem{halpe86_RMP}
W. P. Halperin, 
Quantum size effects in metal particles, 
\href{http://dx.doi.org/10.1103/RevModPhys.58.533}
{Rev. Mod. Phys. \textbf{58}, 533 (1986).}

\bibitem{knigh84_PRL}
W. D. Knight, K. Clemenger, W. A. de Heer, W. A. Saunders, M. Y. Chou, and M. L. Cohen, 
Electronic shell structure and abundances of sodium clusters, 
\href{http://dx.doi.org/10.1103/PhysRevLett.52.2141}
{Phys. Rev. Lett. \textbf{52}, 2141 (1984).}

\bibitem{heer93_RMP}
W. A. de Heer, 
The physics of simple metal clusters: experimental aspects and simple models, 
\href{http://dx.doi.org/10.1103/RevModPhys.65.611}
{Rev. Mod. Phys. \textbf{65}, 611 (1993).}

\bibitem{brack93_RMP}
M. Brack,
The physics of simple metal clusters: self-consistent jellium model and semiclassical approaches, 
\href{http://dx.doi.org/10.1103/RevModPhys.65.677}
{Rev. Mod. Phys. \textbf{65}, 677 (1993).}

\bibitem{cresp04_PRL}
P. Crespo, R. Litr\'an, T. C. Rojas, M. Multigner, J. M. de la Fuente, J. C.
S\'anchez-L\'opez, M. A. Garci\'a, A. Hernando, S. Penad\'es, and A. Fern\'andez, 
Permanent magnetism, magnetic anisotropy, and hysteresis of thiol-capped gold nanoparticles, 
\href{http://dx.doi.org/10.1103/PhysRevLett.93.087204}
{Phys. Rev. Lett. \textbf{93}, 087204 (2004).}

\bibitem{cresp06_PRL}
P. Crespo, M. A. Garc\'ia, E. Fern\'andez Pinel, M. Multigner, D. Alc\'antara, 
J. M. de la Fuente, S. Penad\'es, and A. Hernando, 
Fe impurities weaken the ferromagnetic behavior in Au nanoparticles, 
\href{http://dx.doi.org/10.1103/PhysRevLett.97.177203}
{Phys. Rev. Lett. \textbf{97}, 177203 (2006).}

\bibitem{dutta07_APL}
P. Dutta, S. Pal, M. S. Seehra, M. Anand, and C. B. Roberts, 
Magnetism in dodecanethiol-capped gold nanoparticles: 
Role of size and capping agent, 
\href{https://doi.org/10.1063/1.2740577}
{Appl. Phys. Lett. \textbf{90}, 213102 (2007).}

\bibitem{donni07_AdvMater}
B. Donnio, P. Garc\'ia-V\'azquez, J.-L. Gallani, D. Guillon, and E. Terazzi, 
Dendronized ferromagnetic gold nanoparticles self-organized in a thermotropic 
cubic phase,
\href{http://dx.doi.org/10.1002/adma.200701252}
{Adv. Mater. \textbf{19}, 3534 (2007).}

\bibitem{garit08_NL}
J. S. Garitaonandia, M. Insausti, E. Goikolea, M. Suzuki, J. D. Cashion, 
N. Kawamura, H. Ohsawa, I. Gil de Muro, K. Suzuki, F. Plazaola, and T. Rojo, 
Chemically induced permanent magnetism in Au, Ag, and Cu nanoparticles: 
localization of the magnetism by element selective techniques, 
\href{http://dx.doi.org/10.1021/nl073129g}
{Nano Lett.\ \textbf{8}, 661 (2008).}

\bibitem{guerr08_Nanotechnology}
E. Guerrero, M. A. Mu\~noz-M\'arquez, M. A. Garc\'ia, P. Crespo, 
E. Fern\'andez-Pinel, A. Hernando, and A. Fern\'andez, 
Surface plasmon resonance and magnetism of thiol-capped gold nanoparticles, 
\href{http://dx.doi.org/10.1088/0957-4484/19/17/175701}
{Nanotechnology \textbf{19}, 175701 (2008).}

\bibitem{guerr08}
E. Guerrero, M. A. Mu\~noz-M\'arquez, E. Fern\'andez-Pinel, P. Crespo, 
A. Hernando, and A. Fern\'andez, 
Electronic structure, magnetic properties, and microstructural analysis of 
thiol-functionalized Au nanoparticles: role of chemical and structural 
parameters in the ferromagnetic behaviour, 
\href{http://dx.doi.org/10.1007/s11051-008-9445-5}
{J. Nanopart. Res. \textbf{10}, 179 (2008).}

\bibitem{venta09}
J. de la Venta,  V. Bouzas, A. Pucci, M. A. Laguna-Marco, D. Haskel, 
S. G. E. te Velthuis, A. Hoffmann, J. Lal, M. Bleuel, G. Ruggeri, 
C. de Juli\'an Fern\'andez, and M. A.  Garc\'ia,
X-ray magnetic circular dichroism and small angle neutron scattering studies 
of thiol capped gold nanoparticles,
\href{https://doi.org/10.1166/jnn.2009.1877}
{J. Nanosci. Nanotechnol. \textbf{9}, 6434 (2009).}

\bibitem{donni10_SM}
B. Donnio, A. Derory, E. Terazzi, M. Drillon, D. Guillon, and J.-L. Gallani, 
Very slow high-temperature relaxation of the remnant magnetic moment in 
\unit[2]{nm} mesomorphic gold nanoparticles, 
\href{http://dx.doi.org/10.1039/b918602f}
{Soft Matter \textbf{6}, 965 (2010).}

\bibitem{maitr11_CPC}
U. Maitra, B. Das, N. Kumar, A. Sundaresan, and C. N. R. Rao, 
Ferromagnetism exhibited by nanoparticles of noble metals, 
\href{http://dx.doi.org/10.1002/cphc.201100121}
{ChemPhysChem \textbf{12}, 2322 (2011).}

\bibitem{agrac17_ACSOmega}
M. Agrachev, S. Antonello, T. Dainese, M. Ruzzi, A. Zoleo, E. Apr\`a, N. Govind, 
A. Fortunelli, L. Sementa, and F. Maran, 
Magnetic ordering in gold nanoclusters, 
\href{http://dx.doi.org/10.1021/acsomega.7b00472}
{ACS Omega \textbf{2}, 2607 (2017).}

\bibitem{hori99_JPA}
H. Hori, T. Teranishi, Y. Nakae, Y. Seino, M. Miyake, and S. Yamada, 
Anomalous magnetic polarization effect of Pd and Au nano-particles, 
\href{https://doi.org/10.1016/S0375-9601(99)00742-2}
{Phys. Lett. A \textbf{263}, 406 (1999).}

\bibitem{nakae00_PhysicaB}
Y. Nakae, Y. Seino, T. Teranishi, M. Miyake, S. Yamada, and H. Hori, 
Anomalous spin polarization in Pd and Au nano-particles, 
\href{https://doi.org/10.1016/S0921-4526(99)02928-2}
{Physica B \textbf{284-288}, 1758 (2000).}

\bibitem{hori04_PRB}
H. Hori, Y. Yamamoto, T. Iwamoto, T. Miura, T. Teranishi, and M. Miyake, 
Diameter dependence of ferromagnetic spin moment in Au nanocrystals, 
\href{http://dx.doi.org/10.1103/PhysRevB.69.174411}
{Phys. Rev. B \textbf{69}, 174411 (2004).}

\bibitem{yamam04_PRL}
Y. Yamamoto, T. Miura, M. Suzuki, N. Kawamura, H. Miyagawa, T. Nakamura, 
K. Kobayashi, T. Teranishi, and H. Hori, 
Direct observation of ferromagnetic spin polarization in gold nanoparticles, 
\href{http://dx.doi.org/10.1103/PhysRevLett.93.116801}
{Phys. Rev. Lett. \textbf{93}, 116801 (2004).}

\bibitem{yamam06}
Y. Yamamoto and H. Hori, 
Direct observation of the ferromagnetic spin polarization in gold nanoparticles: 
A review, 
\href{http://www.ipme.ru/e-journals/RAMS/no_11206/yamamoto.html}
{Rev. Adv. Mater. Sci. \textbf{12}, 23 (2006).}

\bibitem{barto12_PRL}
J. Bartolom\'e, F. Bartolom\'e, L. M. Garc\'ia, A. I. Figueroa, A. Repoll\'es, 
M. J. Mart\'inez-P\'erez, F. Luis, C. Mag\'en, S. Selenska-Pobell, F. Pobell, 
T. Reitz, R. Sch\"onemann, T. Herrmannsd\"orfer, M. Merroun,
A. Geissler, F. Wilhelm, and A. Rogalev, 
Strong paramagnetism of gold nanoparticles deposited on a 
sulfolobus acidocaldarius S layer, 
\href{http://dx.doi.org/10.1103/PhysRevLett.109.247203}
{Phys. Rev. Lett. \textbf{109}, 247203 (2012).}

\bibitem{rhee13_PRL}
P. G. van Rhee, P. Zijlstra, T. G. A. Verhagen, J. Aarts, M. I. Katsnelson, 
J. C. Maan, M. Orrit, and P. C. M. Christianen, 
Giant magnetic susceptibility of gold nanorods detected by magnetic alignment, 
\href{http://dx.doi.org/10.1103/PhysRevLett.111.127202}
{Phys. Rev. Lett. \textbf{111}, 127202 (2013).}

\bibitem{garci09_JAP}
M. A. Garcia, E. Fernandez Pinel, J. de la Venta, A. Quesada, V. Bouzas, 
J. F. Fern\'andez, J. J. Romera, M. S. Mart\'in Gonz\'alez, 
and J. L. Costa-Kr\"amer, 
Sources of experimental errors in the observation of nanoscale magnetism, 
\href{http://dx.doi.org/10.1063/1.3060808}
{J. Appl. Phys. \textbf{105}, 013925 (2009).}

\bibitem{nealo12_Nanoscale}
G. L. Nealon, B. Donnio, R. Gr\'eget, J.-P. Kappler, E. Terazzi, and 
J.-L. Gallani, 
Magnetism in gold nanoparticles, 
\href{http://dx.doi.org/10.1039/C2NR30640A}
{Nanoscale \textbf{4}, 5244 (2012).}

\bibitem{donni2017}
B. Donnio, J.-L. Gallani, and M. V. Rastei, 
\textit{Characterization of Magnetism in Gold Nanoparticles}, 
in \href{http://dx.doi.org/10.1007/978-3-662-52780-1}{\textit{Magnetic Characterization Techniques for Nanomaterials}, 
edited by C. S. S. R. Kumar (Springer-Verlag, Berlin, 2017).}

\bibitem{herna06_PRL}
A. Hernando, P. Crespo, and M. A. Garc\'ia, 
Origin of orbital ferromagnetism and giant magnetic anisotropy at the nanoscale, 
\href{http://dx.doi.org/10.1103/PhysRevLett.96.057206}
{Phys. Rev. Lett. \textbf{96}, 057206 (2006).}

\bibitem{imry15_PRB}
Y. Imry, 
Superconducting fluctuations and large diamagnetism of low-$T_\mathrm{c}$ 
nanoparticles, 
\href{http://dx.doi.org/10.1103/PhysRevB.91.104503}
{Phys. Rev. B \textbf{91}, 104503 (2015).}

\bibitem{li11_PRB}
C.-Y. Li, C.-M. Wu, S. K. Karna, C.-W. Wang, D. Hsu, C.-J. Wang, and W.-H. Li, 
Intrinsic magnetic moments of gold nanoparticles, 
\href{http://dx.doi.org/10.1103/PhysRevB.83.174446}
{Phys. Rev. B \textbf{83}, 174446 (2011).}

\bibitem{grege12_CPC}
R. Gr\'eget, G. L. Nealon, B. Vileno, P. Turek, C. M\'eny, F. Ott, A. Derory, 
E. Voirin, E. Rivi\`ere, A. Rogalev, F. Wilhelm, L. Joly, W. Knafo, G. Ballon, 
E. Terazzi, J.-P. Kappler, B. Donnio, and J.-L. Gallani, 
Magnetic properties of gold nanoparticles: a room-temperature quantum effect, 
\href{http://dx.doi.org/10.1002/cphc.201200394}
{ChemPhysChem \textbf{13}, 3092 (2012).}

\bibitem{bohr11_PhD}
N.\ Bohr, PhD thesis, University of Copenhagen, unpublished (1911).

\bibitem{leeuw21_JP}
H.-J. van Leeuwen, 
Probl\`emes de la th\'eorie \'electronique du magn\'etisme, 
\href{http://dx.doi.org/10.1051/jphysrad:01921002012036100}
{J. Phys. Radium \textbf{2}, 361 (1921).}

\bibitem{landa30_ZPhys}
L. D. Landau, 
Diamagnetismus der Metalle, 
\href{http://dx.doi.org/10.1007/BF01397213}
{Z. Phys. \textbf{64}, 629 (1930).}

\bibitem{landau_statphys}
L. D. Landau and E. M. Lifshitz, 
\textit{Statistical Physics}
(Pergamon, Oxford, 1985).

\bibitem{ruite91_PRL}
J. M. van Ruitenbeek and D. A. van Leeuwen, 
Model calculation of size effects in orbital magnetism, 
\href{http://dx.doi.org/10.1103/PhysRevLett.67.640}
{Phys. Rev. Lett. \textbf{67}, 640 (1991).}

\bibitem{ruite93_MPLB}
J. M. van Ruitenbeek and D. A. van Leeuwen, 
Size effects in orbital magnetism, 
\href{https://doi.org/10.1142/S0217984993001053}
{Mod. Phys. Lett. B \textbf{07}, 1053 (1993).}

\bibitem{leuwe93_PhD}
D. A. van Leeuwen, 
\textit{Magnetic Moments in Metacluster Molecules}, 
PhD thesis, University of Leiden, unpublished (1993).

\bibitem{fraue98_PRB}
S. Frauendorf, V. M. Kolomietz, A. G. Magner, and A. I. Sanzhur, 
Supershell structure of magnetic susceptibility, 
\href{http://dx.doi.org/10.1103/PhysRevB.58.5622}
{Phys. Rev. B \textbf{58}, 5622 (1998).}

\bibitem{levy90_PRL}
L. P. L\'evy, G. Dolan, J. Dunsmuir, and H. Bouchiat, 
Magnetization of mesoscopic copper rings: evidence for persistent currents, 
\href{http://dx.doi.org/10.1103/PhysRevLett.64.2074}
{Phys. Rev. Lett. \textbf{64}, 2074 (1990).}

\bibitem{chand91_PRL}
V. Chandrasekhar, R. A. Webb, M. J. Brady, M. B. Ketchen, W. J. Gallagher, 
and A. Kleinsasser, 
Magnetic response of a single, isolated gold loop, 
\href{https://doi.org/10.1103/PhysRevLett.67.3578}
{Phys. Rev. Lett. \textbf{67}, 3578 (1991).}

\bibitem{maill93_PRL}
D. Mailly, C. Chapelier, and A. Benoit, 
Experimental observation of persistent currents in a GaAs-AlGaAs single loop, 
\href{https://doi.org/10.1103/PhysRevLett.70.2020}
{Phys. Rev. Lett. \textbf{70}, 2020 (1993).}

\bibitem{butti83_PLA}
M. B\"uttiker, Y. Imry, and R. Landauer, 
Josephson behavior in small normal one-dimensional rings, 
\href{http://dx.doi.org/10.1016/0375-9601(83)90011-7}
{Phys. Lett. A \textbf{96}, 365 (1983).}

\bibitem{bouch89_JP}
H. Bouchiat and G. Montambaux, 
Persistent currents in mesoscopic rings: 
ensemble averages and half-flux-quantum periodicity, 
\href{http://dx.doi.org/10.1051/jphys:0198900500180269500}
{J. Phys. France \textbf{50}, 2695 (1989).}

\bibitem{imry91}
Y. Imry, in 
\textit{Coherence effects in Condensed Matter Systems}, 
edited by B. Kramer (Plenum, New York, 1991).

\bibitem{schmi91_PRL}
A. Schmid, 
Persistent currents in mesoscopic rings by suppression of charge fluctuations, 
\href{http://dx.doi.org/10.1103/PhysRevLett.66.80}
{Phys. Rev. Lett. \textbf{66}, 80 (1991).}

\bibitem{oppen91_PRL}
F. von Oppen and E. K. Riedel, 
Average persistent current in a mesoscopic ring, 
\href{http://dx.doi.org/10.1103/PhysRevLett.66.84}
{Phys. Rev. Lett. \textbf{66}, 84 (1991).}

\bibitem{altsh91_PRL}
B. L. Altshuler, Y. Gefen, and Y. Imry, 
Persistent differences between canonical and grand canonical averages in 
mesoscopic ensembles: large paramagnetic orbital susceptibilities, 
\href{http://dx.doi.org/10.1103/PhysRevLett.66.88}
{Phys. Rev. Lett. \textbf{66}, 88 (1991).}

\bibitem{mulle93_EPL}
A. M\"uller-Groeling, H. A. Weidenm\"uller, and C. H. Lewenkopf, 
Interacting electrons in mesoscopic rings, 
\href{https://doi.org/10.1209/0295-5075/22/3/006}
{EPL \textbf{22}, 193 (1993).}

\bibitem{blesz09_Science}
A. C. Bleszynski-Jayich, W. E. Shanks, B. Peaudecerf, E. Ginossar, F. von Oppen, 
L. Glazman, and J. G. E. Harris, 
Persistent currents in normal metal rings, 
\href{http://dx.doi.org/10.1126/science.1178139}
{Science \textbf{326}, 272 (2009).}

\bibitem{levy93_PhysicaB}
L. P. L\'evy, D. H. Reich, L. Pfeiffer, and K. West, 
Aharonov-Bohm ballistic billiards, 
\href{http://dx.doi.org/10.1016/0921-4526(93)90161-X}
{Physica B \textbf{189}, 204 (1993).}

\bibitem{oppen94_PRB}
F. von Oppen, 
Magnetic susceptibility of ballistic microstructures, 
\href{http://dx.doi.org/10.1103/PhysRevB.50.17151}
{Phys. Rev. B \textbf{50}, 17151 (1994).}

\bibitem{ullmo95_PRL}
D. Ullmo, K. Richter, and R. A. Jalabert, 
Orbital magnetism in ensembles of ballistic billiards, 
\href{http://dx.doi.org/10.1103/PhysRevLett.74.383}
{Phys. Rev. Lett. \textbf{74}, 383 (1995).}

\bibitem{richt96_PhysRep}
K. Richter, D. Ullmo, and R. A. Jalabert, 
Orbital magnetism in the ballistic regime: geometrical effects, 
\href{http://dx.doi.org/10.1016/0370-1573(96)00010-5}
{Phys. Rep. \textbf{276}, 1 (1996).}

\bibitem{richt98_EPL}
K. Richter and B. Mehlig, 
Orbital magnetism of classically chaotic quantum systems, 
\href{https://doi.org/10.1209/epl/i1998-00197-2}
{EPL \textbf{41}, 587 (1998).}

\bibitem{richt96_PRB}
K. Richter, D. Ullmo, and R. A. Jalabert, 
Smooth-disorder effects in ballistic microstructures, 
\href{http://dx.doi.org/10.1103/PhysRevB.54.R5219}
{Phys. Rev. B \textbf{54}, R5219 (1996).}

\bibitem{richt96_JMP}
K. Richter, D. Ullmo, and R. A. Jalabert, 
Integrability and disorder in mesoscopic systems: 
application to orbital magnetism, 
\href{http://dx.doi.org/10.1063/1.531677}
{J. Math. Phys. \textbf{37}, 5087 (1996).}

\bibitem{ullmo98_PRL}
D. Ullmo, H. U. Baranger, K. Richter, F. von Oppen, and R. A. Jalabert, 
Chaos and interacting electrons in ballistic quantum dots, 
\href{http://dx.doi.org/10.1103/PhysRevLett.80.895}
{Phys. Rev. Lett. \textbf{80}, 895 (1998).}

\bibitem{gutzwiller}
M. C. Gutzwiller, 
\textit{Chaos in Classical and Quantum Mechanics} 
(Springer-Verlag, Berlin, 1990).

\bibitem{brack}
M. Brack and R. K. Bhaduri, 
\textit{Semiclassical Physics}
(Addison-Wesley, Reading, 1997).

\bibitem{range12_PRB}
T. Rangel, D. Kecik, P. E. Trevisanutto, G.-M. Rignanese, H. Van Swygenhoven,
and V. Olevano, 
Band structure of gold from many-body perturbation theory, 
\href{http://dx.doi.org/10.1103/PhysRevB.86.125125}
{Phys. Rev. B \textbf{86}, 125125 (2012).}

\bibitem{mathu91_PRB}
H. Mathur and A. D. Stone, 
Persistent-current paramagnetism and spin-orbit interaction in mesoscopic rings, 
\href{http://dx.doi.org/10.1103/PhysRevB.44.10957}
{Phys. Rev. B \textbf{44}, 10957(R) (1991).}

\bibitem{grege12_thesis}
R. Gr\'eget, \textit{Propri\'et\'es magn\'etiques de nanoparticules d'or 
fonctionnalis\'ees}, PhD thesis, Universit\'e de Strasbourg, unpublished (2012).

\bibitem{weick05_PRB}
G. Weick, R. A. Molina, D. Weinmann, and R. A. Jalabert, 
Lifetime of the first and second collective excitations in metallic 
nanoparticles, 
\href{http://dx.doi.org/10.1103/PhysRevB.72.115410}
{Phys. Rev. B \textbf{72}, 115410 (2005).}

\bibitem{weick06_PRB}
G. Weick, G.-L. Ingold, R. A. Jalabert, and D. Weinmann, 
Surface plasmon in metallic nanoparticles: 
renormalization effects due to electron-hole excitations, 
\href{http://dx.doi.org/10.1103/PhysRevB.74.165421}
{Phys. Rev. B \textbf{74}, 165421 (2006).}

\bibitem{tanak96_PRB}
K. Tanaka, S. C. Creagh, and M. Brack, 
Simple metal clusters in magnetic fields, 
\href{http://dx.doi.org/10.1103/PhysRevB.53.16050}
{Phys. Rev. B \textbf{53}, 16050 (1996).}

\bibitem{weick11_PRB}
G. Weick and D. Weinmann, 
Lifetime of the surface magnetoplasmons in metallic nanoparticles, 
\href{http://dx.doi.org/10.1103/PhysRevB.83.125405}
{Phys. Rev. B \textbf{83}, 125405 (2011).}

\bibitem{footnote:T_F}
This statement is, strictly speaking, valid for temperatures much smaller than
the Fermi temperature. 
As we are dealing with metals, this is fulfilled in all experimentally-relevant
situations.

\bibitem{footnote:Au}
The cyclotron radius $R_\mathrm{c}=\frac{7.6}{B}\mathrm{G\cdot cm}$ for gold.
Other relevant parameters for gold are the Fermi energy
$E_\mathrm{F}=\unit[5.5]{eV}$, the Fermi temperature
$T_\mathrm{F}=\unit[6.4\times10^4]{K}$, and the Fermi wave vector
$k_\mathrm{F}=\unit[1.2\times10^{8}]{cm^{-1}}$.

\bibitem{footnote:chi_cgs}
In cgs units, the (dimensionless) magnetic susceptibility $\chi$ is linked 
to the susceptibility in SI units $\chi_\mathrm{SI}$ through the relation 
$\chi_\mathrm{SI}=4\pi\chi$.

\bibitem{gutzw70_JMP}
M. C. Gutzwiller, 
Energy spectrum according to classical mechanics, 
\href{http://dx.doi.org/10.1063/1.1665328}
{J. Math. Phys. \textbf{11}, 1791 (1970).}

\bibitem{gutzw71_JMP}
M. C. Gutzwiller, 
Periodic orbits and classical quantization conditions, 
\href{http://dx.doi.org/10.1063/1.1665596}
{J. Math. Phys. \textbf{12}, 343 (1971).}

\bibitem{berry76_PRSL}
M. V. Berry and M. Tabor, 
Closed orbits and the regular bound spectrum, 
\href{http://dx.doi.org/10.1098/rspa.1976.0062}
{Proc. R. Soc. Lond. A \textbf{349}, 101 (1976).}

\bibitem{berry77_JPA}
M. V. Berry and M. Tabor, 
Calculating the bound spectrum by path summation in action-angle variables, 
\href{http://dx.doi.org/10.1088/0305-4470/10/3/009}
{J. Phys. A \textbf{10}, 371 (1977).}

\bibitem{footnote:DOS}
In Eq.\ \eqref{eq:rho_osc}, we neglect the contribution of the diametral
orbits $(1,2)$ as it is of higher order in $\hbar$ 
(see Ref.\ \cite{tanak96_PRB}). 
Moreover, diametral trajectories yield, for small fields, a field-independent 
contribution to the density of states that does not contribute to the 
magnetization.

\bibitem{saenz87_JAP}
J. J. S\'aenz, N. Garc\'ia, P. Gr\"utter, E. Meyer, H. Heinzelmann, 
R. Wiesendanger, L. Rosenthaler, H. R. Hidber, and H.-J. G\"untherodt, 
Observation of magnetic forces by the atomic force microscope, 
\href{http://dx.doi.org/10.1063/1.339105}
{J. Appl. Phys. \textbf{62}, 4293 (1987).}

\bibitem{footnote:nondiag}
Given the high numerical cost of performing the four summations of 
Eq.\ \eqref{eq:M_ens} at the low temperature of the experiment, we limited 
ourselves to the case $\nu'=\nu$ (verifying that for the corresponding value 
of $k_\mathrm{F}\delta a$ the exponential factor strongly suppresses the terms 
having $\nu'\neq\nu$), and we replaced the thermal factor $R_T$ by a Heaviside 
step function, similarly to what we have done in Secs.\ \ref{sec:1NP} 
and \ref{sec:manyNPs} when deriving Eqs.\ \eqref{eq:chi_1_SC} and \eqref{eq:C}, 
respectively (verifying also the applicability of such an approximation).

\bibitem{yamam_privatecomm}
Y. Yamamoto (private communication). 

\bibitem{hikam80}
S. Hikami, A. I. Larkin, and Y. Nagaoka, 
Spin-orbit interaction and magnetoresistance in the two dimensional 
random system, 
\href{http://dx.doi.org/10.1143/PTP.63.707}
{Prog. Theor. Phys. \textbf{63}, 707 (1980).}

\bibitem{mezard}
M. Mezard, G. Parisi, and M. A. Virasoro, 
\textit{Spin Glass Theory and Beyond} 
(World Scientific, Singapore, 1987).

\bibitem{fischer}
K. A. Fischer and J. A. Hertz, 
\textit{Spin Glasses} 
(Cambridge University Press, Cambridge, 1991).

\end{thebibliography}
\end{document}